\documentclass[lettersize,journal]{IEEEtran}
\usepackage{amsmath,amsfonts} % 数学公式库
\usepackage{algorithm,algorithmic}
\usepackage{array} % 阵列库
\usepackage{textcomp} % 特殊符号库
\usepackage{stfloats} % 浮动体相关库
\usepackage{url} % 链接库
\usepackage{color} % 颜色库
\usepackage{verbatim} % 原样抄写库
\usepackage{graphicx} % 图片库
\usepackage{epstopdf}
\usepackage{cite} % 引用库
\usepackage{multicol} % 多栏单栏包
\usepackage{float}
\usepackage{amssymb}
\usepackage{subfigure}
\def\BibTeX{{\rm B\kern-.05em{\sc i\kern-.025em b}\kern-.08em
		T\kern-.1667em\lower.7ex\hbox{E}\kern-.125emX}}
\usepackage{balance}
\newcounter{TempEqCnt}
\newcommand*{\RMN}[1]{\uppercase\expandafter{\romannumeral#1}}
\begin{document}
	%% 标题
	\title{Unified Design of Space-Air-Ground-Sea Integrated Maritime Communications}
	\author{\IEEEauthorblockN{
			\normalsize Zhehan~Zhou, Xiaoming~Chen, Ming~Ying, Zhaohui~Yang, Chongwen~Huang, Yunlong~Cai, and~Zhaoyang~Zhang}
			
	\thanks{Zhehan Zhou, Xiaoming Chen, Ming Ying, Zhaohui Yang, Chongwen Huang, Yunlong Cai, and Zhaoyang Zhang are with the College of Information Science and Electronic Engineering, Zhejiang University, Hangzhou 310027, China (e-mail: \{3200100787, chen\_xiaoming, ming\_ying, yang\_zhaohui, chongwenhuang, ylcai, zhzy\}@zju.edu.cn).}
	}\maketitle
	%{Shell \MakeLowercase{\textit{et al.}}: A Sample Article Using IEEEtran.cls for IEEE Journals}% 不能注释，否则出错
	
	%\IEEEpubid{0000--0000/00\$00.00~\copyright~2021 IEEE} % 脚注
	
	%% 摘要
	\begin{abstract}
		With the explosive growth of maritime activities, it is expected to provide seamless communications with quality of service (QoS) guarantee over broad sea area. In the context, this paper proposes a space-air-ground-sea integrated maritime communication architecture combining satellite, unmanned aerial vehicle (UAV), terrestrial base station (TBS) and unmanned surface vessel (USV). Firstly, according to the distance away from the shore, the whole marine space is divided to coastal area, offshore area, middle-sea area and open-sea area,  the maritime users in which are served by TBS, USV, UAV and satellite, respectively. Then,  by exploiting the potential of integrated maritime communication system, a joint beamforming and trajectory optimization algorithm is designed to maximize the minimum transmission rate of maritime users. Finally, theoretical analysis and simulation results validate the effectiveness of the proposed algorithm.
	\end{abstract}
	%% 关键词
	\begin{IEEEkeywords}
		Maritime communication, integrated space-air-ground-sea, beamforming, trajectory design.
	\end{IEEEkeywords}
	
	\section{INTRODUCTION}
	\IEEEPARstart{C}{urrently}, the ever-growing human activities on the ocean has promoted enormous demand for high-throughput maritime communication services \cite{background9453860,Survey,background}. In contrast to terrestrial scenario, harsh environment and immense serving area with short of facilities make an urgent demand for a dedicated maritime communication architecture, which poses great challenge to the conventional maritime communication network (MCN) \cite{ChanelApp9344715}.
	Therefore, hybrid satellite-terrestrial MCN, which integrates both maritime satellite and shore-based network, has aroused great interest \cite{satellite_terrestrial_integration,Coordinated_satellite_terrestrial_network}.
	The framework combines the advantages of satellite's wide coverage with sophisticated technology of terrestrial base station (TBS). % revision
	However, the high overhead and scarce spectrum resource of satellite communication limit its application for ordinary maritime users.
	
	To provide affordable service and reduce the dependency on the satellite, multi-hop network has been extensively researched, which exploits island-based, air-based, vessel-based relays and so on to extend the coverage \cite{island_relay,tower_relay,ballon_relay,buoy_relay}. Yet, numerous studies on maritime multi-hop networks tend to focus on static relays or vessels restricted to fixed trajectories. Such structures lack the capability of dynamically adapting to the ever-changing maritime environment and network topology. Therefore, it is vital to establish a more flexible deployment structure by integrating adjustable relays into the hybrid MCNs.
	
	To this end, unmanned aerial vehicle (UAV) is becoming a promising solution for agile communication,
	which has been widely researched in satellite-terrestrial-UAV integrated MCNs \cite{UAV,UAVkinetic8770592,NOMA_uav,UAV_delay_analysis}. Moreover, a multi-UAV coordinated network was illustrated in \cite{multi_UAV}. Commonly, UAV exhibits distinct advantage in adjusting spatial location tailored for on-demand communication in maritime applications. However, UAV is subject to limited load for battery and hostile weather condition on the ocean.
	Thus, several attempts have been made to employ unmanned surface vessel (USV) in maritime scenario \cite{USVkinetic9547273,VIChannel9956800}.
	In \cite{USVkinetic9547273}, a USV was deployed to assist on-demand service for maritime users. The authors in \cite{VIChannel9956800} utilized multiple-input multiple-output technology in a USV-enabled system to achieve performance enhancement. Compared with UAV, USV displays superior payload and endurance, which enables USV to equip advanced communication devices \cite{USV_research}. Nevertheless, the stability of USV communication service is greatly impaired by frequent fluctuation of sea surface.
	Furthermore, deep reinforcement learning-based solutions have been investigated to achieve trajectory planning without sufficient prior knowledge \cite{DRLisacUAV,DRLmultiUAV,DRLVHN}.
	Overall, the aforementioned MCNs are unable to provide reliable service of low cost and wide coverage, since their deployments are not complete and unable to fully conquer the difficulties in maritime communication. Thus, it makes sense to integrate space-air-ground-sea devices in a unified architecture and resolves the critical challenge of resources and users allocation among inter-influenced links.
	
	Despite of the aforementioned works, there lacks researches that focus on the complete evaluation of the space-air-ground-sea integrated MCN. For instance, reference \cite{uav_usv_operation} specified the collaborative efforts for the UAVs and USVs in maritime communication yet did not offer the numerical model and evaluation for the deployment. In \cite{NOMA_uav}, the authors modeled a hybrid satellite-UAV-terrestrial network and formulated a throughput maximization algorithm for UAV links, but the performance enhancement for satellite and TBS links was not yet considered in the algorithm. To the best of the authors' knowledge, the research on the systemic evaluation of the entire space-air-ground-sea integrated MCN remains an open issue.
	
	Motivated by the constraints of current MCNs and the limitations of related researches, this paper considers a space-air-ground-sea integrated maritime communication system and investigates the cooperative strategies of different devices. We develop a unique scheduling strategy that assigns users to optimal service areas with optimal communication devices. Specifically, USV and UAV are deployed as mobile relays to expand the coverage and four different links exploiting advantages of TBS, USV, UAV and satellite are respectively assigned to deal with different demands. The major contributions of our work are listed as follows.
	
	{\begin{itemize}
			\item A unified space-air-ground-sea hierarchical architecture is designed for seamless maritime communication. The marine space is divided to four areas and arranged four communication modes combining the advantages of TBS, USV, UAV and satellite to realize on-demand service.
			\item A feasible algorithm is proposed by jointly optimizing trajectory and beamforming. It uniquely develops a unified scheduling strategy that dynamically assigns users to optimal service areas with optimal communication modes and resolves the critical challenge of resources and users allocation among four inter-influenced links.
			\item Extensive simulation results demonstrate that the proposed algorithm outperforms other methods in the fairness of resource allocation, adaption to diverse scenarios and ability to mitigate adjacent-channel interference.
	\end{itemize}}
	
	The rest of this paper is organized as follows. In Section \RMN{2}, we introduce the space-air-ground-sea integrated architecture for maritime communications. Section \RMN{3} formulates the optimization problem and develops a joint beamforming and trajectory optimization algorithm. The convergence and complexity are analyzed. Then, we conduct simulations to assess the performance of the proposed algorithm in Section \RMN{4}. Finally, the paper is concluded in Section \RMN{5}.
	
	\textit{Notations}: Boldface upper letters and boldface lower letters indicate the matrices and vectors, while scalars are represented by italic letters. $\mathbb{C}^{M \times N }$ denotes the space of complex matrices of size $M \times N$. $(\cdot)^T$, $(\cdot)^*$ and $(\cdot)^H$ denote the transpose, conjugation and conjugate transpose, respectively. $\lvert \cdot \rvert$ and $\lVert \cdot \rVert$ denote the absolute value and 2-norm, respectively. $\mathcal{N} (\mu,\sigma^2)$ indicates the Gaussian distribution with mean $\mu$ and variance $\sigma^2$. For a complex value $x$, $\text{Re}\{x\}$ means the real part of $x$.
	%\ref{table}.
	%% Problem Formulation
	\section{SYSTEM MODEL}
\begin{figure}[t]
		\centering
		\includegraphics[width=0.8\columnwidth]{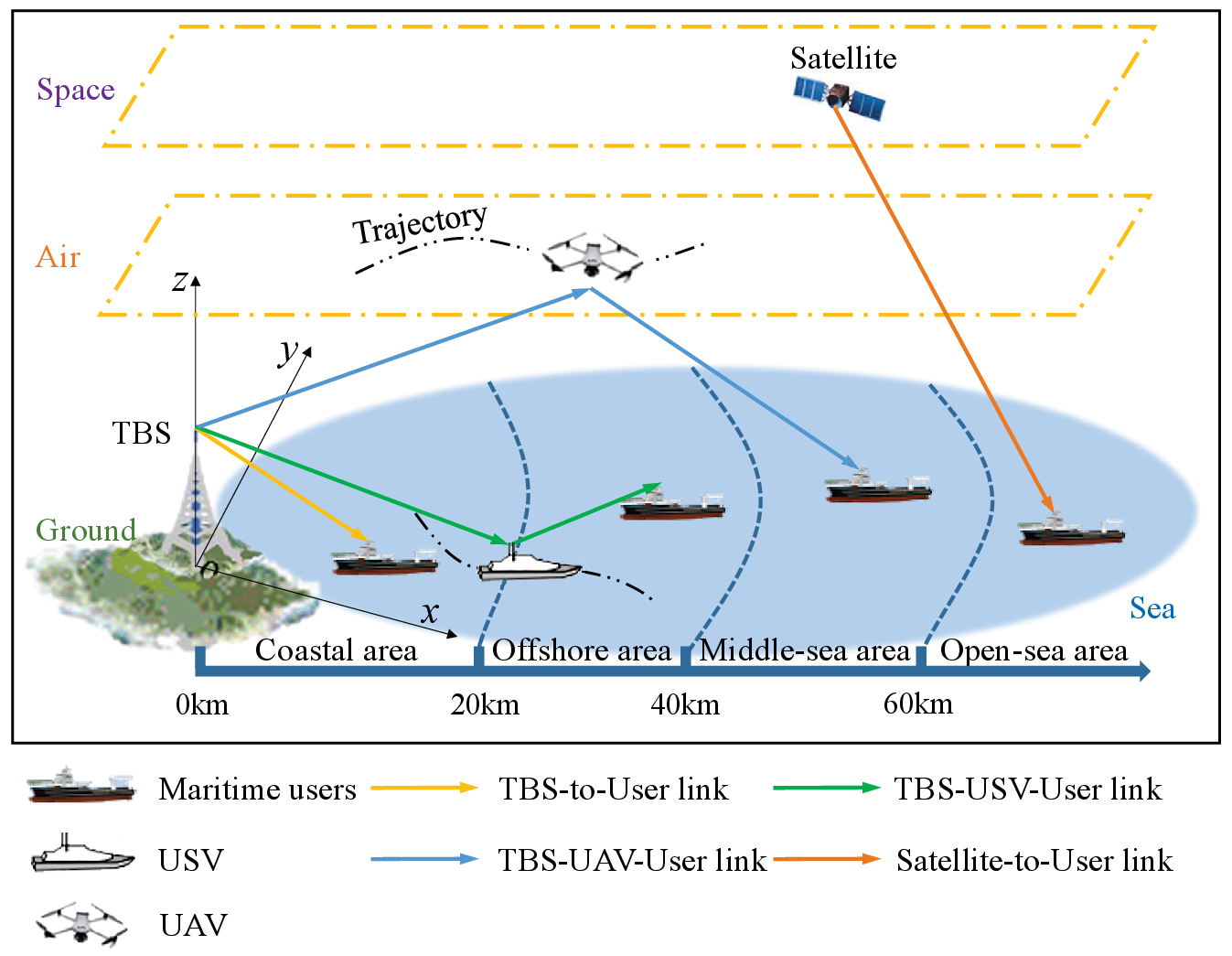}
		\caption{Space-air-ground-sea integrated maritime communication system model.}
		\label{SystemImg}
	\end{figure}

	Considering a space-air-ground-sea integrated maritime communication system as shown in Fig. \ref{SystemImg}. The whole marine space is divided into four areas according to the distance away from the shore, namely coastal area, offshore area, middle-sea area, and open-sea area. An on-shore TBS equipped with $N_b$ antennas provides services to $M_b$ TBS maritime users in the coastal area, while an LEO satellite equipped with $N_s$ antennas transmits signals to $M_s$ satellite maritime users in the open-sea area. Due to limited wireless coverage of the TBS, a USV equipped with $N_v$ antennas and a UAV equipped with $N_a$ antennas are deployed as mobile relays to provide on-demand communication services from the TBS to $M_v$ USV maritime users and $M_a$ UAV maritime users in the offshore area and middle-sea area, respectively.\footnote{We assume that, with elaborate design, multiple USVs and UAVs are sparsely distributed on the immense ocean to avoid mutual interference within each other. Thus, without loss of generality, we take only one USV and one UAV into consideration, similar assumption in other related work \cite{footnote1}.}
	The user sets of TBS, USV, UAV and satellite are denoted as $\mathcal{BI}$, $\mathcal{VI}$, $\mathcal{AI}$ and $\mathcal{SI}$, respectively. For ease of exposition, we define the maritime user set $i \in \mathcal{I} \triangleq \{bi \in \mathcal{BI}, vi \in \mathcal{VI}, ai \in \mathcal{AI}, si \in \mathcal{SI}\}$, where $bi$, $vi$, $ai$ and $si$ denote the $i$-th maritime user of the TBS, USV, UAV and satellite, respectively. We suppose that all maritime users are equipped with a single antenna.
	
	The maritime communication system is established on a three-dimensional coordinate system, where the TBS is located at $\mathbf{q}_b=[0,0]^T$ with fixed antenna height $h_b$. The transmission period $T$ is discretized into $N$ time slots, i.e., $T=N\Delta \tau$. We assume that $\Delta \tau$ is sufficiently small such that the locations and channel state information (CSI) can be viewed as unchanged in each time slot. Thus the continuous trajectories of USV, UAV and maritime user $i$ are transformed into coordinate sequences as $\{ (x_v[n], y_v[n], h_v)^T\}_{n=1}^N$, $\{ (x_a[n], y_a[n], h_a)^T\}_{n=1}^N$ and $\{ (x_i[n], y_i[n], h_i)^T\}_{n=1}^N$. It is reasonably assumed that the antenna heights of USV, UAV and maritime user $i$, i.e., $h_v$, $h_a$ and $h_i$, keep constant during finite transmission period, and thus the signal strength fluctuation caused by antenna height variation is negligible \cite{height_negligible}.
	
	In the following contents, we provide channel models and signal models in the considered space-air-ground-sea integrated maritime communications.
			%% a long-bottom equation must be put on the previous page
	%\newcounter{TempEqCnt} % 创建临时变量TempEqCnt
	
	\subsection{Channel Model}
	In the proposed space-air-ground-sea integrated maritime communication system, there are four communication links, namely TBS-to-User link for coastal communication, TBS-USV-User link for offshore communication, TBS-UAV-User link for middle-sea communication, and satellite-to-User link for open-sea communication. Herein, we introduce these links in detail.
	%%  TBS-To-User Link
	\subsubsection{TBS-to-User link}
	\
	\newline
	\indent In the coastal area, the TBS is able to provide high-quality services directly to maritime users. In general, maritime channel is mainly dominated by the line-of-sight (LOS) path and sea-surface-reflected ray \cite{VIChannel8528349}. Considering the typical two-ray channel model with multi-reflection paths from rough sea surface \cite{VIChanneltest}, the channel model from TBS to maritime user $i$ can be expressed as
	\begin{equation}
		\mathbf{h}_{b,i} \! = \! \frac{\lambda\!\sin\!\left(\!\frac{2\pi h_b h_i}{\lambda d_{b,i}}\!\right)}{2\pi d_{b,i}} \! \left(\!\sqrt{\frac{K_{b,i}}{K_{b,i}+1}}\! \mathbf{h}_{b,i}^{\mathrm{LOS}}\!+\!\sqrt{\frac{1}{K_{b,i}+1}}\! \mathbf{h}_{b,i}^{\mathrm{NLOS}}\!\right), \label{h_bi}
	\end{equation}
	with $\lambda$ being the wavelength of carrier wave, $d_{b,i}$ being the link distance from TBS to maritime user $i$ as
	\begin{equation}
		d_{b,i}=\sqrt{\lVert\mathbf{q}_b-\mathbf{q}_i \rVert^2+(h_b-h_i)^2},
	\end{equation}
	where $\mathbf{q}_{i}=[x_i, y_i]^T$.
	
	For small-scale fading, $K_{b,i}$ denotes the Rician factor, $\mathbf{h}_{b,i}^{\mathrm{LOS}}\in \mathbb{C}^{N_b \times 1}$ denotes the LOS component and presents as antenna array response (AAR) in mathematical form according to the Saleh-Valenzuela Model, which is given by \cite{AIChannel8974403}
	\begin{equation}
		\mathbf{h}_{b,i}^{\mathrm{LOS}}=\left(1, \dotsb, e^{-\bar{b}(n_b-1)\cos \vartheta_{b,i}}, \dotsb , e^{-\bar{b}(N_b-1)\cos \vartheta_{b,i}}\right)^T,
		\label{AARbi}
	\end{equation}
	where $\bar{b}=j2 \pi b/\lambda$ with $b$ being the antenna spacing, $\vartheta_{b,i}$ denotes the angle-of-departure (AoD) of the TBS-to-User link. $\mathbf{h}_{b,i}^{\mathrm{NLOS}}\in \mathbb{C}^{N_b \times 1}$ denotes the non-line-of-sight (NLOS) component, following the independent and identically distributed (i.i.d.) complex Gaussian distribution.	
	%% TBS-USV-User Link
	\subsubsection{TBS-USV-User link}
	\
	\newline
	\indent In the offshore area, the communications from the TBS should be aided by a USV due to long transmission distance. Specifically, a USV sets out from initial point $\mathbf{q}_v[0]$ and travels along the trajectory $\{ \mathbf{q}_v[n] \}_{n=1}^N=\{ (x_v[n], y_v[n])^T\}_{n=1}^N$ to assist on-demand communication service. During the transmission period, the USV receives desired signal sent from the TBS and relays it to the target user, thus the TBS-USV-User link contains both signal reception and forwarding at the USV.\footnote{To facilitate the analysis, we assume that the USV and UAV utilize self-interference cancellation to suppress the self-interference to the noise floor \cite{SIC,Self_interference}.}
	
	Two-ray channel model is also applicable to TBS-USV-User link. Specifically, the maritime channels from TBS to USV and USV to maritime user $i$ can be respectively expressed as
	\begin{align}
		\mathbf{H}_{b,v}&\!\!=\!\!\frac{\lambda\!\sin\!\left(\!\frac{2\pi h_b h_v}{\lambda d_{b,v}}\!\right)}{2\pi d_{b,v}}\!\! \left(\!\!\sqrt{\frac{K_{b,v}}{K_{b,v}+1}}\! \mathbf{H}_{b,v}^{\mathrm{LOS}}\!\!+\!\!\sqrt{\frac{1}{K_{b,v}+1}}\! \mathbf{H}_{b,v}^{\mathrm{NLOS}}\!\!\right), \label{h_bv}\\
		\mathbf{h}_{v,i}&\!\!=\!\!\frac{\lambda\sin\left(\frac{2\pi h_v h_i}{\lambda d_{v,i}}\right)}{2\pi d_{v,i}}\!\! \left(\!\!\sqrt{\frac{K_{v,i}}{K_{v,i}+1}}\! \mathbf{h}_{v,i}^{\mathrm{LOS}}\!\!+\!\!\sqrt{\frac{1}{K_{v,i}+1}}\! \mathbf{h}_{v,i}^{\mathrm{NLOS}}\!\!\right),  \label{h_vi}
	\end{align}
	with $d_{b,v}$ and $d_{v,i}$ being the link distances from TBS to USV and USV to maritime user $i$, which are given by
	\begin{align}
		d_{b,v}&=\sqrt{\lVert\mathbf{q}_b-\mathbf{q}_v \rVert^2+(h_b-h_v)^2},\\
		d_{v,i}&=\sqrt{\lVert\mathbf{q}_v-\mathbf{q}_i \rVert^2+(h_v-h_i)^2}.
	\end{align}
	
	For small-scale fading of respective USV channels, $K_{b,v}$ and $K_{v,i}$ denote the Rician factors, $\mathbf{H}_{b,v}^{\mathrm{LOS}}\in \mathbb{C}^{N_b \times N_v}$ and $\mathbf{h}_{v,i}^{\mathrm{LOS}}\in \mathbb{C}^{N_v \times 1}$ denote the LOS components given by
	\begin{equation}
		\begin{aligned}
			\label{AARbv}\mathbf{H}_{b,v}^{\mathrm{LOS}}&=\left(1, \dotsb, e^{-\bar{b}(n_b-1)\cos \vartheta_{b,v}}, \dotsb, e^{-\bar{b}(N_b-1)\cos \vartheta_{b,v}}\right)^T \\
			&\times\left(1, \dotsb, e^{-\bar{b}(n_v-1)\cos \theta_{b,v}}, \dotsb, e^{-\bar{b}(N_v-1)\cos \theta_{b,v}}\right)^*,
		\end{aligned}
	\end{equation}
	\begin{equation}
		\mathbf{h}_{v,i}^{\mathrm{LOS}}=\left(1, \dotsb, e^{-\bar{b}(n_v-1)\cos \vartheta_{v,i}}, \dotsb, e^{-\bar{b}(N_v-1)\cos \vartheta_{v,i}}\right)^T,
		\label{AARvi}
	\end{equation}
	where $\theta_{b,v}$ and $\vartheta_{b,v}$ denote the angle-of-arrival (AoA) and AoD of the TBS-to-USV link, and $\vartheta_{v,i}$ denotes the AoD of the USV-to-User link.
	Finally, $\mathbf{H}_{b,v}^{\mathrm{NLOS}}\in \mathbb{C}^{N_b \times N_v}$ and $\mathbf{h}_{v,i}^{\mathrm{NLOS}}\in \mathbb{C}^{N_v \times 1}$ denote the NLOS components which follow the i.i.d. complex Gaussian distribution.
	
	Generally, the USV is restricted to its own maximum speed $V_{v}^{\mathrm{max}}$. Then, the USV should obey the mobility constraint as below \cite{USVkinetic9547273}
	\begin{equation}
		\lVert\mathbf{q}_v[n]-\mathbf{q}_v[n-1] -\Delta \tau \mathbf{v}_c[n]\rVert \leq \Delta \tau V_{v}^{\max}, \label{MaxUSVSpC.}
	\end{equation}
	where $\mathbf{v}_c$ denotes the ocean current velocity.
	
	To avoid rolling or even rollover, the steering angle of the USV is usually limited to a maximum value $\varphi_v^{\mathrm{max}}$ \cite{KIM201437}. We can describe the steering angle constraint of USV as \cite{USVkinetic9547273}
	\begin{equation}
		\frac{\left( \mathbf{q}_v[n]-\mathbf{q}_v[n-1] \right)^T \left( \mathbf{q}_v[n-1]-\mathbf{q}_v[n-2] \right)}{\lVert  \mathbf{q}_v[n]-\mathbf{q}_v[n-1]  \rVert \lVert  \mathbf{q}_v[n-1]-\mathbf{q}_v[n-2]  \rVert} \geq \cos \varphi_v^{\max},\label{USVangleC.}
	\end{equation}
	as for the steering angle constraint at time slot $n=1$, we introduce a base vector $\mathbf{x}$ and equivalently transform the constraint (\ref{USVangleC.}) into
	\begin{equation}
		\frac{\left( \mathbf{q}_v[1]-\mathbf{q}_v[0] \right)^T \mathbf{x}}{\lVert  \mathbf{q}_v[1]-\mathbf{q}_v[0]  \rVert \lVert \mathbf{x} \rVert}
		\geq \cos \varphi_v^{\max}, \label{USVangleInitC.}
	\end{equation}
	where $\mathbf{x}$ is the unit vector of $x$-axis.
	
	We further suppose that there are $K$ obstacles, such as reefs or buoys, distributed in the serving area and their locations are denoted as $\mathbf{o}_k=[x_k,y_k]^T$. To prevent collision, the USV needs to keep a safe distance from obstacles and maritime users, namely $r_{\mathrm{ob},k}$ and $r_{\mathrm{sh}}$. Thus, the USV should obey the following safe sailing constraints \cite{USVkinetic9547273}
	\begin{align}
		\label{USVsafeObC.}\lVert \mathbf{q}_v[n]-\mathbf{o}_k[n] \rVert &\geq r_{\mathrm{ob},k}, \forall k, n,\\
		\label{USVsafeShC.}\lVert \mathbf{q}_v[n]-\mathbf{q}_i[n] \rVert &\geq r_{\mathrm{sh}}, \forall i, n.
	\end{align}
	%% Maritime Air-To-Sea Channel
	\subsubsection{TBS-UAV-User link}
	\
	\newline
	\indent The UAV is employed as a mobile relay with faster speed than the USV, which enables the UAV to travel longer distance to provide services for maritime users in the middle-sea area. Specifically, a UAV sets off from $\mathbf{q}_a[0]$ and flies along the trajectory $\{\mathbf{q}_a[n]\}_{n=1}^N=\{ (x_a[n], y_a[n])^T\}_{n=1}^N$ to bridge the TBS-UAV-User link. Similarly, the TBS-UAV-User link consists of both signal reception and forwarding at the UAV.
	
	Generally, the UAV flies at a sufficiently high altitude to enable LOS transmission. The maritime channels from TBS to UAV and UAV to maritime user $i$ can be modeled as \cite{footnote1}
	\begin{align}
		{\mathbf{G}_{b,a}}&=\frac{\rho}{p_{b,a}}\left(\sqrt{\frac{K_{b,a}}{K_{b,a}+1}} {\mathbf{G}_{b,a}^{\mathrm{LOS}}}+\sqrt{\frac{1}{K_{b,a}+1}} {\mathbf{G}_{b,a}^{\mathrm{NLOS}}}\right), \label{h_ba} \\
		{\mathbf{g}_{a,i}}&=\frac{\rho}{p_{a,i}}\left(\sqrt{\frac{K_{a,i}}{K_{a,i}+1}} {\mathbf{g}_{a,i}^{\mathrm{LOS}}}+\sqrt{\frac{1}{K_{a,i}+1}} {\mathbf{g}_{a,i}^{\mathrm{NLOS}}}\right), \label{h_ai}
	\end{align}
	with $\rho$ being the channel gain at unit reference distance, $p_{b,a}$ and $p_{a,i}$ being the link distances from TBS to UAV and UAV to maritime user $i$, which can be computed as
	\begin{align}
		&p_{b,a}=\sqrt{\lVert\mathbf{q}_b-\mathbf{q}_a \rVert^2+(h_b-h_a)^2},\\
		&p_{a,i}=\sqrt{\lVert\mathbf{q}_a-\mathbf{q}_i \rVert^2+(h_a-h_i)^2}.
	\end{align}
	
	For small-scale fading of respective UAV channels, $K_{b,a}$ and $K_{a,i}$ denote the Rician factors, ${\mathbf{G}_{b,a}^{\mathrm{LOS}}} \in \mathbb{C}^{N_b \times N_a}$ and ${\mathbf{g}_{a,i}^{\mathrm{LOS}}} \in \mathbb{C}^{N_a \times 1}$ denote the LOS components, which are given by
	\begin{equation}
		\begin{aligned}
			\label{AARba}{\mathbf{G}_{b,a}^{\mathrm{LOS}}}&=\left(1, \dotsb, e^{-\bar{b}(n_b-1)\cos \vartheta_{b,a}}, \dotsb, e^{-\bar{b}(N_b-1)\cos \vartheta_{b,a}}\right)^T \\
			&\times\left(1, \dotsb, e^{-\bar{b}(n_a-1)\cos \theta_{b,a}}, \dotsb, e^{-\bar{b}(N_a-1)\cos \theta_{b,a}}\right)^*,
		\end{aligned}
	\end{equation}
	\begin{equation}
		{\mathbf{g}_{a,i}^{\mathrm{LOS}}}=\left(1, \dotsb, e^{-\bar{b}(n_a-1)\cos \vartheta_{a,i}}, \dotsb, e^{-\bar{b}(N_a-1)\cos \vartheta_{a,i}}\right)^T,
		\label{AARai}
	\end{equation}
	where $\theta_{b,a}$ and $\vartheta_{b,a}$ denote the AoA and AoD of the TBS-to-UAV link, and $\vartheta_{a,i}$ denotes the AoD of the UAV-to-User link. ${\mathbf{G}_{b,a}^{\mathrm{NLOS}}} \in \mathbb{C}^{N_b \times N_a}$ and ${\mathbf{g}_{a,i}^{\mathrm{NLOS}}} \in \mathbb{C}^{N_a \times 1}$ denote the NLOS components which follow the i.i.d. complex Gaussian distribution.
	
	Similar to USV, UAV is subject to the mobility constraints and steering angle constraints, which are expressed as \cite{UAVkinetic8770592}
	\begin{equation}
		\lVert\mathbf{q}_a[n]-\mathbf{q}_a[n-1] -\Delta \tau \mathbf{v}_w[n]\rVert \leq \Delta \tau V_{a}^{\max},
		\label{MaxUAVSpC.}
	\end{equation}
	\begin{equation}
		\frac{\left( \mathbf{q}_a[n]-\mathbf{q}_a[n-1] \right)^T \left( \mathbf{q}_a[n-1]-\mathbf{q}_a[n-2] \right)}{\lVert  \mathbf{q}_a[n]-\mathbf{q}_a[n-1]  \rVert \lVert  \mathbf{q}_a[n-1]-\mathbf{q}_a[n-2]  \rVert} \geq \cos \varphi_a^{\max},
		\label{UAVangleC.}
	\end{equation}
	\begin{equation}
		\frac{\left( \mathbf{q}_a[1]-\mathbf{q}_a[0] \right)^T \mathbf{x}}{\lVert  \mathbf{q}_a[1]-\mathbf{q}_a[0]  \rVert \lVert \mathbf{x} \rVert}
		\geq \cos \varphi_a^{\max}, \label{UAVangleInitC.}
	\end{equation}
	where $\mathbf{v}_w$ denotes the wind velocity. Since there is hardly any aerial obstacle at the altitude of UAV, the safe flying constraint is not considered.
	%% Satellite-To-User Link
	\subsubsection{Satellite-to-User link}
	\
	\newline
	\indent In the open-sea area, the LEO satellite is utilized to provide effective communication services. According to the signal propagation features of space-to-sea communication, maritime satellite channel is mainly affected by factors of pathloss, atmospheric impairment, and antenna gain \cite{SIChannel8654189}. Thus, the space-to-sea channel between satellite and maritime user $i$ is modeled as \cite{SIChannel}
	\begin{equation}
		{\mathbf{f}_{s,i}} \! = \! \frac{\lambda}{4\pi h_{s}}\!\sqrt{\frac{G_i\omega_{s,i}}{\beta_{s,i}}}\! \left(\!\sqrt{\frac{K_{s,i}}{K_{s,i}+1}}\!{\mathbf{f}_{s,i}^{\mathrm{LOS}}}\!+\!\sqrt{\frac{1}{K_{s,i}+1}}\!{\mathbf{f}_{s,i}^{\mathrm{NLOS}}}\!\right),
		\label{SIChannel1}
	\end{equation}
	where $\frac{\lambda}{4\pi h_s}$ represents the free space loss with $h_s$ being the height of satellite, $G_i$ is the antenna gain of maritime user $i$ and $\beta_{s,i}$ is the rain attenuation coefficient. $\omega_{s,i}$ indicates the satellite antenna gain given by \cite{SIChannel9343295}
	\begin{equation}
		\omega_{s,i} = \left[\omega_{i}^{\max} \left( \frac{J_1(\phi_{s,i})}{2\phi_{s,i}}+36\frac{J_3(\phi_{s,i})}{\phi_{s,i}^3} \right)\right]^2,
		\label{SIChannel3}
	\end{equation}
	where $\omega_{i}^{\max}$ denotes the maximum antenna gain of LEO satellite beam, $\phi_{s,i} = \frac{\pi l_s }{\lambda}\sin(\theta_{s,i})$ with $l_s$ being the diameter of satellite's circular antenna array, $J_1$ and $J_3$ being the first and third order of the first-kind Bessel
	function, respectively.
	
	For small-scale fading, $K_{s,i}$ denotes the Rician factor, ${\mathbf{f}_{s,i}^{\mathrm{LOS}}} \in \mathbb{C}^{N_s \times 1}$ denotes the LOS component and ${\mathbf{f}_{s,i}^{\mathrm{NLOS}}} \in \mathbb{C}^{N_s \times 1}$ denotes the NLOS component following the i.i.d. complex Gaussian distribution.
	
	Based on the above channel models in space-air-ground-sea integrated maritime communications, we discuss the signal models in what follows.
	\subsection{Signal Model}
	\begin{figure}[H]
		\centering
		\includegraphics[width=0.8\columnwidth]{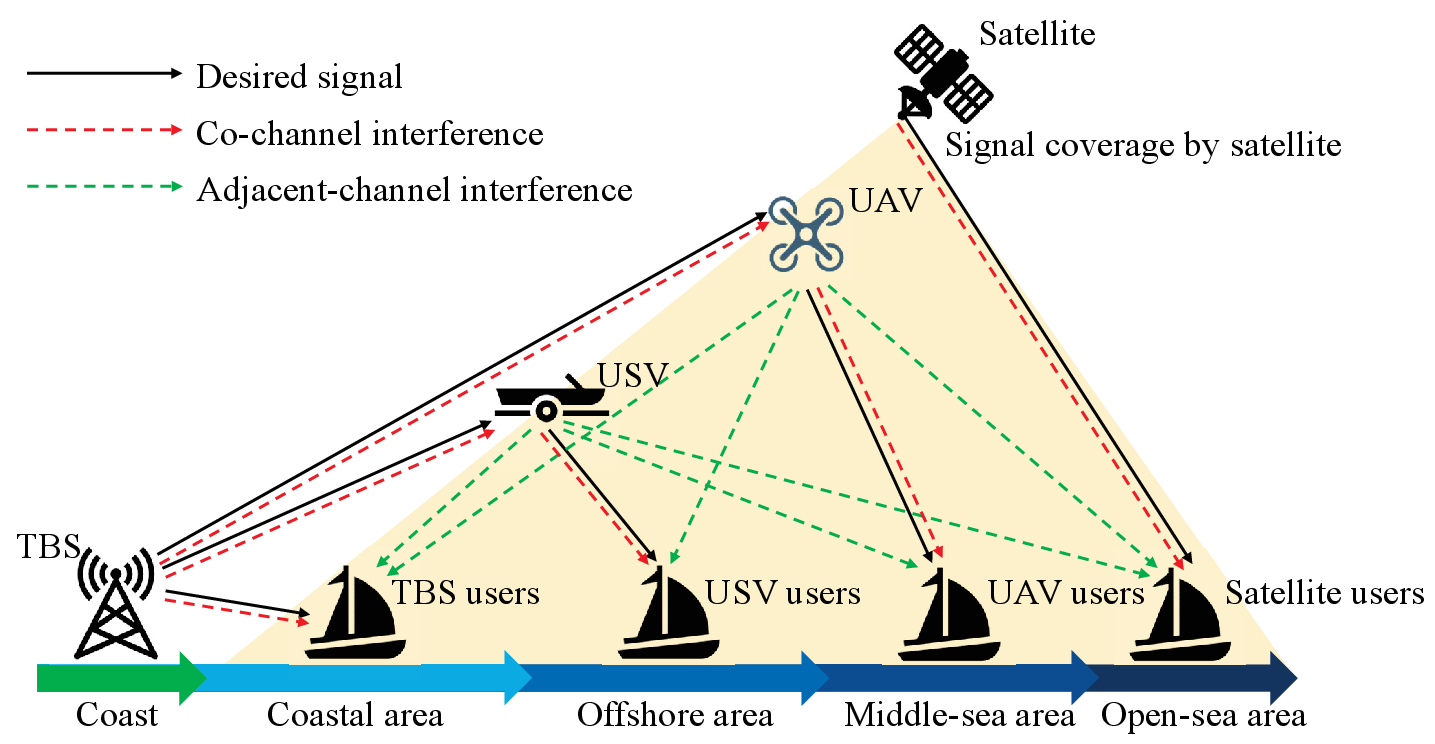}
		\caption{Desired signal and interference distribution of the proposed maritime communication system.}
		\label{SignalImg}
	\end{figure}
	To solve the spectrum scarcity problem and support high-throughput service, the entire maritime communication system works in the shared 2 GHz carrier frequency \cite{VHF,ITU}.
	As a result, there exists complicated interference among the four groups of users, as shown in Fig. \ref{SignalImg}.
	
	To be specific, the signal from the TBS to maritime users far away from the coast is negligible considering the strong long-distance signal attenuation in maritime environment. On the other hand, due to the wide coverage of the satellite and high maneuverability of the USV and UAV, the four groups of users are generally interfered by these devices.
	Moreover, the Doppler shift is found to be negligible for maritime wireless communications and can be pre-compensated in the receiver \cite{Doppler,Compensate}. Following previous relevant work \cite{Doppler7572068}, it is reasonably assumed that the Doppler shift in the considered system model is compensated perfectly in advance.
	In this case, the received signals of the four groups of users are as follows.
	\subsubsection{Received Signal of TBS User}
	\
	\newline
	\indent The TBS users receive signals sent from the TBS, USV, UAV and satellite. Hence, the received signal of TBS user $bi$ at time slot $n$ is given by
	\begin{equation}
		\begin{aligned}
			y_{bi} &= \underbrace{\mathbf{h}_{b,bi}^H\mathbf{w}_{bi}x_{bi}}_{\text{Desired signal}} \!+\!
			 \underbrace{\mathbf{h}_{b,bi}^H\!\left(\!\sum_{j=1,j\neq i}^{M_b}\!\!\!\mathbf{w}_{bj}x_{bj} \!+\! \mathbf{w}_{bv}x_{bv} \!+\! \mathbf{w}_{ba}x_{ba}\!\right)}_{\text{Co-channel interference }}\\
			 &+\underbrace{\mathbf{h}_{v,bi}^H\sum_{j=1}^{M_v}\mathbf{w}_{vj}x_{vj}+
			{\mathbf{g}_{a,bi}^H} \sum_{j=1}^{M_a}\mathbf{w}_{aj}x_{aj} + {\mathbf{f}_{s,bi}^H}\sum_{j=1}^{M_s}\mathbf{w}_{sj}x_{sj}}_{\text{Adjacent-channel interference}}\\
			&+\underbrace{n_{bi}}_{\text{AWGN}}, \label{y_bi}
		\end{aligned}
	\end{equation}
	where $\mathbf{h}_{b,bi}$, $\mathbf{h}_{v,bi}$, ${\mathbf{g}_{a,bi}}$, and ${\mathbf{f}_{s,bi}}$ denote the channels from the TBS, USV, UAV and satellite to TBS user $bi$,
	$\mathbf{w}_{bj}\in \mathbb{C}^{N_b \times 1}$, $\mathbf{w}_{bv}\in \mathbb{C}^{N_b \times 1}$ and $\mathbf{w}_{ba}\in \mathbb{C}^{N_b \times 1}$ represent the beamforming vectors utilized for transmission from the TBS to TBS users $bj$, USV and UAV, respectively,
	$\mathbf{w}_{aj}\in \mathbb{C}^{N_a \times 1}$, $\mathbf{w}_{vj}\in \mathbb{C}^{N_v \times 1}$ and $\mathbf{w}_{sj}\in \mathbb{C}^{N_s \times 1}$ represent the beamforming vectors used at the UAV, USV and satellite,
	$x_{bj}$, $x_{bv}$ and $x_{ba}$ denote the transmitted signal from the TBS to TBS user $bj$, USV and UAV,
	$x_{aj}$, $x_{vj}$ and $x_{sj}$ denote the transmitted signal from the UAV, USV and satellite to respective target user,
	and $n_{bi}$ denotes the additive white Gaussian noise (AWGN) with zero mean and variance $\sigma_{bi}^2$.
	
	Based on the received signal in (\ref{y_bi}), the transmission rate associated with the TBS user $bi$ at time slot $n$ can be computed as (\ref{R_bi}).
	% insert R_bv,R_vi expression, shown at next page
	\setcounter{TempEqCnt}{\value{equation}}
	\begin{figure*}[hb]
		\hrulefill
		\begin{equation}
			\begin{aligned}
				R_{bi} = \log_2\left(1+\frac{\lvert \mathbf{h}_{b,bi}^H\mathbf{w}_{bi} \rvert^2} {\sum\limits_{j=1,j \neq i}^{M_b}\!\!\!\lvert \mathbf{h}_{b,bi}^H\mathbf{w}_{bj} \rvert^2 +
					\lvert \mathbf{h}_{b,bi}^H\mathbf{w}_{bv} \rvert^2 +
					\lvert \mathbf{h}_{b,bi}^H\mathbf{w}_{ba} \rvert^2 + \sum\limits_{j=1}^{M_v}\lvert \mathbf{h}_{v,bi}^H\mathbf{w}_{vj} \rvert^2 +
					\sum\limits_{j=1}^{M_a}\lvert {\mathbf{g}_{a,bi}^H}\mathbf{w}_{aj} \rvert^2 +
					\sum\limits_{j=1}^{M_s}\lvert {\mathbf{f}_{s,bi}^H}\mathbf{w}_{sj} \rvert^2 + \sigma_{bi}^2}\right),
			\end{aligned}
			\label{R_bi}
		\end{equation}
		\setcounter{equation}{30}
		\begin{equation}
			\begin{aligned}
				R_{vi}=\log_{2}\left( 1 +
				\frac{\lvert \mathbf{h}_{v,vi}^H\mathbf{w}_{vi} \rvert^2}
				{\sum_{j=1,j \neq i}^{M_v}\lvert \mathbf{h}_{v,vi}^H\mathbf{w}_{vj} \rvert^2 +
					\sum_{j=1}^{M_a}\lvert {\mathbf{g}_{a,vi}^H}\mathbf{w}_{aj} \rvert^2 + \sum_{j=1}^{M_s}\lvert {\mathbf{f}_{s,vi}^H}\mathbf{w}_{sj} \rvert^2 + \sigma_{vi}^2} \right),
			\end{aligned}
			\label{R_vi}
		\end{equation}
	\end{figure*}
	\subsubsection{Received Signal of USV User}
	\
	\newline
	\indent Firstly, the USV receives the signal sent by the TBS. In this way, the received signal of the USV and corresponding transmission rate at time slot $n$ are given by (\ref{y_bv}) and (\ref{R_bv}), respectively
	\footnote{The focus of the work lies in optimizing the transmit-side beamforming scheme to enhance the transmit rate (\ref{R_bv}), (\ref{R_ba}). While, the receiver is simplified to adopt well-established detection techniques, such as MMSE or ZF beamforming \cite{Detection}.}
	\setcounter{equation}{27}
	\begin{equation}
		\begin{aligned}
			{\mathbf{y}_{bv}}&=\underbrace{\mathbf{H}_{b,v}^H \mathbf{w}_{bv} x_{bv}}_{\text{Desired signal}}+\underbrace{\mathbf{H}_{b,v}^H \left(\mathbf{w}_{ba}x_{ba}+\sum_{j=1}^{M_b}\mathbf{w}_{bj} x_{bj} \right)}_{\text{Co-channel interference}}+\underbrace{{ \mathbf{n}_{bv}}}_{\text{AWGN}},
			\label{y_bv}
		\end{aligned}
	\end{equation}
	\begin{equation}
		\begin{aligned}
			R_{bv}=\log_{2}\left( 1+
			\frac{\lVert \mathbf{H}_{b,v}^H\mathbf{w}_{bv} \rVert^2}
			{\lVert \mathbf{H}_{b,v}^H\mathbf{w}_{ba} \rVert^2+
				\sum_{j=1}^{M_b}\lVert \mathbf{H}_{b,v}^H\mathbf{w}_{bj} \rVert^2+\sigma_{bv}^2} \right),
			\label{R_bv}
		\end{aligned}
	\end{equation}
	where $\mathbf{n}_{bv}\in \mathbb{C}^{N_v \times 1}$ is the AWGN vector with $\mathbb{E}\left[\mathbf{n}_{bv}\mathbf{n}^{H}_{bv}\right] = \sigma_{bv}^2 \mathbf{I}_{N_v}$.
	
	Once receiving the signal from TBS, the USV decodes it, obtains the desired signal and then forwards to each USV users. The received signal of USV user $vi$ and its transmission rate at time slot $n$ are given by (\ref{y_vi}) and (\ref{R_vi}), respectively
	\begin{equation}
		\begin{aligned}
			y_{vi} &= \underbrace{\mathbf{h}_{v,vi}^H \mathbf{w}_{vi}x_{vi}}_{\text{Desired signal}}+\underbrace{\mathbf{h}_{v,vi}^H \!\!\!\sum_{j=1,j \neq i}^{M_v}\!\!\!\mathbf{w}_{vj}x_{vj}}_{\text{Co-channel interference}}\\
			&+\underbrace{{\mathbf{g}_{a,vi}^H}\sum_{j=1}^{M_a}\mathbf{w}_{aj}x_{aj}+{\mathbf{f}_{s,vi}^H}\sum_{j=1}^{M_s}\mathbf{w}_{sj}x_{sj}}_{\text{Adjacent-channel interference}} +\underbrace{n_{vi}}_{\text{AWGN}},
			\label{y_vi}
		\end{aligned}
	\end{equation}
	where $\mathbf{h}_{v,vi}$, ${\mathbf{g}_{a,vi}}$, ${\mathbf{f}_{s,vi}}$ denote the channels from the USV, UAV and satellite to USV user $vi$, $n_{vi}$ denotes the AWGN with zero mean and variance $\sigma_{vi}^2$.

Since the USV can only forward the signal from the TBS, the sum rate of all USV users cannot exceed that from TBS to USV. Hence, we have the following information-causality constraint \cite{VIChannel9956800}
	\setcounter{equation}{31}
	\begin{align}
		\label{USVCausalC.}\sum_{j=1}^{M_v} R_{vj} \leq R_{bv}.
	\end{align}
	
	\subsubsection{Received Signal of UAV User}
	\
	\newline
	\indent	Similar to USV, the received signal of UAV, the received signal of UAV user $ai$ and their transmission rates at time slot $n$ are given by (\ref{y_ba}), (\ref{y_ai}), (\ref{R_ba}) and (\ref{R_ai}), respectively
	\begin{equation}
		\label{y_ba}
		{\mathbf{y}_{ba}}=\underbrace{{\mathbf{G}_{b,a}^H}\mathbf{w}_{ba}x_{ba}}_{\text{Desired signal}}+ \underbrace{{\mathbf{G}_{b,a}^H}\left( \mathbf{w}_{bv}x_{bv} + \sum_{j=1}^{M_b}\mathbf{w}_{bj}x_{bj}\right)}_{\text{Co-channel interference}} +\underbrace{{\mathbf{n}_{ba}}}_{\text{AWGN}}
	\end{equation}
	\begin{align}
		\label{y_ai}
		y_{ai}&=\underbrace{{\mathbf{g}_{a,ai}^H}\mathbf{w}_{ai}x_{ai}}_{\text{Desired signal}}+\underbrace{{\mathbf{g}_{a,ai}^H}\!\!\!\sum_{j=1,j \neq i}^{M_a}\!\!\!\mathbf{w}_{aj}x_{aj}}_{\text{Co-channel interference}}\nonumber\\
		&+\underbrace{\mathbf{h}_{v,ai}^H\sum_{j=1}^{M_v}\mathbf{w}_{vj}x_{vj}+{\mathbf{f}_{s,ai}^H}\sum_{j=1}^{M_s}\mathbf{w}_{sj}x_{sj}}_{\text{Adjacent-channel interference}}+\underbrace{n_{ai}}_{\text{AWGN}},
	\end{align}
	\begin{equation}
		\begin{aligned}
			R_{ba}=\log_{2}\left(1+
			\frac{\lVert {\mathbf{G}_{b,a}^H}\mathbf{w}_{ba} \rVert^2}
			{\lVert {\mathbf{G}_{b,a}^H}\mathbf{w}_{bv} \rVert^2+
				\sum_{j=1}^{M_b}\lVert {\mathbf{G}_{b,a}^H}\mathbf{w}_{bj} \rVert^2+\sigma_{ba}^2} \right),
			\label{R_ba}
		\end{aligned}
	\end{equation}
	where $\mathbf{g}_{a,ai}$, $\mathbf{h}_{v,ai}$, ${\mathbf{f}_{s,ai}}$ denote the channels from the UAV, USV and satellite to UAV user $ai$, respectively, $n_{ai}$ denotes the AWGN with zero mean and variance $\sigma_{ai}^2$, and $\mathbf{n}_{ba}\in \mathbb{C}^{N_a \times 1}$ is the AWGN vector with $\mathbb{E}\left[\mathbf{n}_{ba}\mathbf{n}^{H}_{ba}\right] = \sigma_{ba}^2 \mathbf{I}_{N_a}$.
	
	Additionally, the information-causality constraint is similarly imposed on the TBS-UAV-User link, which can be expressed as
	\setcounter{equation}{36}
	\begin{align}
		\label{UAVCausalC.}\sum_{j=1}^{M_a} R_{aj} \leq R_{ba}.
	\end{align}
	\subsubsection{Received Signal of Satellite User}
	\
	\newline
	\indent	The satellite users receive signals from the satellite, USV and UAV. Then, the received signal of satellite user $si$ and its transmission rate at time slot $n$ are given by (\ref{y_si}) and (\ref{R_si})
	\begin{equation}
		\begin{aligned}
			y_{si} &= \underbrace{{\mathbf{f}_{s,si}^H}\mathbf{w}_{si}x_{si}}_{\text{Desired signal}}+\underbrace{{\mathbf{f}_{s,si}^H}\!\!\!\sum_{j=1,j \neq i}^{M_s}\!\!\!\mathbf{w}_{sj}x_{sj}}_{\text{Co-channel interference}}\\
			&+\underbrace{{\mathbf{g}_{a,si}^H}\sum_{j=1}^{M_a}\mathbf{w}_{aj}x_{aj}+\mathbf{h}_{v,si}^H\sum_{j=1}^{M_v}\mathbf{w}_{vj}x_{vj}}_{\text{Adjacent-channel interference}}+\underbrace{n_{si}}_{\text{AWGN}},
			\label{y_si}
		\end{aligned}
	\end{equation}
	where ${\mathbf{f}_{s,si}}$, ${\mathbf{g}_{a,si}}$, $\mathbf{h}_{v,si}$ denote the channels from the satellite, UAV and USV to satellite user $si$, $n_{si}$ denotes the AWGN with zero mean and variance $\sigma_{si}^2$.
	\setcounter{TempEqCnt}{\value{equation}}
	\begin{figure*}[hb]
		\hrulefill
		\setcounter{equation}{35}
		\begin{equation}
			\begin{aligned}
				R_{ai}=\log_{2}\left(1+
				\frac{\lvert {\mathbf{g}_{a,ai}^H}\mathbf{w}_{ai} \rvert^2}
				{\sum_{j=1,j \neq i}^{M_a}\lvert {\mathbf{g}_{a,ai}^H}\mathbf{w}_{aj} \rvert^2 +
					\sum_{j=1}^{M_v}\lvert \mathbf{h}_{v,ai}^H\mathbf{w}_{vj} \rvert^2 +
					\sum_{j=1}^{M_s} \lvert {\mathbf{f}_{s,ai}^H}\mathbf{w}_{sj} \rvert^2 + \sigma_{ai}^2} \right),
			\end{aligned}
			\label{R_ai}
		\end{equation}
		\setcounter{equation}{38}
		\begin{equation}
			R_{si}= \log_{2}\left(1+
			\frac{\lvert {\mathbf{f}_{s,si}^H}\mathbf{w}_{si} \rvert^2}
			{\sum_{j=1,j \neq i}^{M_s}\lvert {\mathbf{f}_{s,si}^H}\mathbf{w}_{sj} \rvert^2 + \sum_{j=1}^{M_a}\lvert {\mathbf{g}_{a,si}^H}\mathbf{w}_{aj} \rvert^2 + \sum_{j=1}^{M_v}\lvert \mathbf{h}_{v,si}^H\mathbf{w}_{vj} \rvert^2 + \sigma_{si}^2} \right),
			\label{R_si}
		\end{equation}
	\end{figure*}
	\setcounter{equation}{44} % 当前公式序号变为x，x等于长公式应有的序号减1.
	\setcounter{equation}{\value{TempEqCnt}}
	
	It is seen that the transmission rates of four groups of maritime users are jointly determined by beamforming and trajectory. Hence, it makes sense to give a unified design for the space-air-ground-sea maritime communications.

	\section{Unified Design of Maritime Communication}
	In this section, we first formulate an optimization problem for the unified design of maritime communications, then develop a joint beamforming and trajectory optimization algorithm, finally analyze the effectiveness of the algorithm.
	\subsection{Problem Formulation}
	 Let $\mathcal{Q}=\{ \mathbf{q}_a,\mathbf{q}_v\}$ indicate the trajectory and $\mathcal{W}=\{\mathbf{w}_{bv},\\\mathbf{w}_{ba},\mathbf{w}_{bi},\mathbf{w}_{vi},\mathbf{w}_{ai},\mathbf{w}_{si},\forall bi, vi, ai, si\}$ indicate the beamforming. To ensure performance fairness, our objective is to maximize the minimum transmission rate of all users by jointly optimizing $\mathcal{W}$ and $\mathcal{Q}$. Thus, the optimization problem is formulated as
	 	\setcounter{equation}{39}
	 \begin{align}
	 	\max_{\mathcal{Q},\mathcal{W}}& \ \eta\qquad\qquad\qquad\qquad\qquad\qquad\label{opProb}\\
	 	\textrm{s.t.}& \mathbf{q}_v[0]=[x_0,y_0]^T,\mathbf{q}_a[0]=[x_0,y_0]^T, \tag{\ref{opProb}{a}}\\
	 	&\sum_{j=1}^{M_b}\lVert \mathbf{w}_{bj} \rVert^2 + \lVert \mathbf{w}_{bv} \rVert^2 + \lVert \mathbf{w}_{ba}\rVert^2 \leq P_b ,\tag{\ref{opProb}{b}}\\
	 	&\sum_{j=1}^{M_v} \lVert \mathbf{w}_{vj} \rVert^2 \leq P_v, \tag{\ref{opProb}{c}}\\
	 	&\sum_{j=1}^{M_a} \lVert \mathbf{w}_{aj} \rVert^2 \leq P_a, \tag{\ref{opProb}{d}}\\
	 	&\sum_{j=1}^{M_s} \lVert \mathbf{w}_{sj} \rVert^2 \leq P_s, \tag{\ref{opProb}{e}}\\
	 	&\eta\leq \min\limits_{i}\{R_{bi},R_{vi},R_{ai},R_{si}\}, \tag{\ref{opProb}{f}}\\
	 	&(\ref{MaxUSVSpC.})-(\ref{USVsafeShC.}), (\ref{MaxUAVSpC.})-(\ref{UAVangleInitC.}), (\ref{USVCausalC.}),(\ref{UAVCausalC.}).\nonumber
	\end{align}
	Specifically, (\ref{opProb}{a}) indicates the initial location constraint, (\ref{opProb}{b})-(\ref{opProb}{e}) represent the transmit power constraints with $P_b$, $P_v$, $P_a$, $P_s$ being the maximum transmit power of the TBS, USV, UAV and satellite, and (\ref{opProb}{f}) indicates that the objective function $\eta$ denotes the minimum transmission rate of all users.
	
	For problem (\ref{opProb}), it is difficult to be solved directly for the following reasons. Firstly, the expressions for transmission rates are very complex, particularly the two-ray channel models which contain a sine function.
	Furthermore, numerous non-convex constraints exist in the problem, which cannot be handled directly.
	Additionally, two variable sets $\mathcal{Q}$ and $\mathcal{W}$ are coupled, making the problem quite complicated.
	
	\subsection{Algorithm Design}
	Firstly, we tackle the two-ray channel models in TBS-to-User link and TBS-USV-User link. Since the maritime users tend to scatter dispersedly on the vast sea, we suppose that the link distance is much longer than the antenna height. Thus, we can approximate the two-ray channel models as \cite{USVkinetic9547273}
	\begin{align}
		\mathbf{h}_{b,i} \approx\frac{h_bh_i}{d_{b,i}^2}\! \left(\!\sqrt{\frac{K_{b,i}}{K_{b,i}+1}} \mathbf{h}_{b,i}^{\mathrm{LOS}}\!+\!\sqrt{\frac{1}{K_{b,i}+1}} \mathbf{h}_{b,i}^{\mathrm{NLOS}}\! \right), \label{Apph_bi}
	\end{align}
	\begin{align}
		\mathbf{h}_{b,v}\! \approx\! \frac{h_bh_v}{d_{b,v}^2}\! \left(\!\sqrt{\frac{K_{b,v}}{K_{b,v}+1}} \mathbf{h}_{b,v}^{\mathrm{LOS}}\!+\!\sqrt{\frac{1}{K_{b,v}+1}} \mathbf{h}_{b,v}^{\mathrm{NLOS}}\! \right), \label{Apph_bv}
	\end{align}
	 \begin{align}
	 	\mathbf{h}_{v,i} \approx\frac{h_vh_i}{d_{v,i}^2}\! \left(\!\sqrt{\frac{K_{v,i}}{K_{v,i}+1}} \mathbf{h}_{v,i}^{\mathrm{LOS}}\!+\!\sqrt{\frac{1}{K_{v,i}+1}} \mathbf{h}_{v,i}^{\mathrm{NLOS}}\! \right), \label{Apph_vi}
	 \end{align}
	which enables us to effectively eliminate the intractable sine functions.
	
	Then, we handle the information-causality constraints (\ref{USVCausalC.}) and (\ref{UAVCausalC.}). Let $\mathcal{R}=\{ r_{vi} \triangleq R_{vi},  r_{ai} \triangleq R_{ai},\forall vi, ai\}$ be the slack variable set of transmission rates, and then the problem can be rewritten as
	\begin{align}
		\max_{\mathcal{Q},\mathcal{W},\mathcal{R}}& \ \eta \label{opProb1}\\
		\textrm{s.t.}&\sum_{j=1}^{M_a} r_{aj} \leq R_{ba}, \tag{\ref{opProb1}{a}}\\
		&\sum_{j=1}^{M_v} r_{vj} \leq R_{bv}, \tag{\ref{opProb1}{b}}\\
		&\eta \leq r_{ai} \leq R_{ai}, \forall ai \tag{\ref{opProb1}{c}}\\
		&\eta \leq r_{vi} \leq R_{vi}, \forall vi \tag{\ref{opProb1}{d}}\\
		&\eta \leq R_{bi}, \forall bi \tag{\ref{opProb1}{e}}\\
		&\eta \leq R_{si}, \forall si \tag{\ref{opProb1}{f}}\\
		&(\ref{MaxUSVSpC.})-(\ref{USVsafeShC.}), (\ref{MaxUAVSpC.})-(\ref{UAVangleInitC.}), (\ref{opProb}{a})-(\ref{opProb}{e}).\nonumber
	\end{align}
	
	To solve the problem (\ref{opProb1}), we adopt the alternative optimization method. In other words, problem (\ref{opProb1}) is decomposed into two sub-problems, namely beamforming optimization with fixed trajectory and trajectory optimization with given beamforming. In each time slot, two sub-problems are iteratively optimized until the objective function converges.
	\setcounter{TempEqCnt}{\value{equation}}
	\setcounter{equation}{44}
	\begin{figure*}[hb]
		\hrulefill
		\begin{equation}
			\begin{aligned}
				R_{si}=\log_{2}\left(\sum_{j = i}^{M_s}\lvert {\mathbf{f}_{s,si}^H}\mathbf{w}_{sj} \rvert^2 + \sum_{j=1}^{M_a}\lvert {\mathbf{g}_{a,si}^H}\mathbf{w}_{aj} \rvert^2 + \sum_{j=1}^{M_v}\lvert \mathbf{h}_{v,si}^H\mathbf{w}_{vj} \rvert^2 + \sigma_{si}^2 \right)-\tilde{R}_{si},
				\label{diffR_si}
			\end{aligned}
		\end{equation}
		where $\tilde{R}_{si}$ is given by
		\begin{equation}
			\tilde{R}_{si} = \log_{2}\left(\sum_{j=1,j \neq i}^{M_s}\!\!\!\lvert {\mathbf{f}_{s,si}^H}\mathbf{w}_{sj} \rvert^2 +
			\sum_{j=1}^{M_a}\lvert {\mathbf{g}_{a,si}^H}\mathbf{w}_{aj} \rvert^2 +
			\sum_{j=1}^{M_v}\lvert \mathbf{h}_{v,si}^H\mathbf{w}_{vj} \rvert^2 + \sigma_{si}^2 \right),
		\end{equation}
	\end{figure*}
	\setcounter{equation}{\value{TempEqCnt}}
	\subsubsection{Beamforming Optimization with Fixed Trajectory}
	\
	\newline
	\indent In this sub-problem, we optimize beamforming $\mathcal{W}$ with given trajectory $\mathcal{Q}$. In this case, (\ref{opProb1}{a})-(\ref{opProb1}{f}) are non-convex constraints with respect to $\mathcal{W}$. Firstly, we rewrite $R_{si}$ into the difference of two logarithmic functions as shown in (\ref{diffR_si}). Then, the difference of convex functions (DC) programming technique is utilized by approximating $\tilde{R}_{si}$ with the first-order Taylor expansion as its upper-bound at the given point. For convenience, let
	$T_{si}=\sum_{j=1,j \neq i}^{M_s}\lvert{\mathbf{f}_{s,si}^H}\mathbf{w}_{sj} \rvert^2 +
	\sum_{j=1}^{M_a}\lvert {\mathbf{g}_{a,si}^H}\mathbf{w}_{aj} \rvert^2 +
	\sum_{j=1}^{M_v}\lvert \mathbf{h}_{v,si}^H\mathbf{w}_{vj} \rvert^2$ and we can obtain
	\setcounter{equation}{46}
	\begin{equation}
		\begin{aligned}
			\tilde{R}_{si} &\leq \frac{\log_{2}\left(e\right)}{T_{si}^{(l)}+\sigma_{si}^2}\! \left(T_{si}-T_{si}^{(l)} \right)\! +\! \log_{2}\!\left(T_{si}^{(l)} + \sigma_{si}^2\right)\!\triangleq\! \tilde{R}_{si}^{\mathrm{up}},
		\end{aligned}
		\label{Rup_si}
	\end{equation}
	where $T_{si}^{(l)}=\sum_{j=1,j \neq i}^{M_s}\lvert{\mathbf{f}_{s,si}^H}\mathbf{w}_{sj}^{(l)} \rvert^2 +
	\sum_{j=1}^{M_a}\lvert {\mathbf{g}_{a,si}^H}\mathbf{w}_{aj}^{(l)} \rvert^2 +
	\sum_{j=1}^{M_v}\lvert \mathbf{h}_{v,si}^H\mathbf{w}_{vj}^{(l)} \rvert^2$ with $\mathbf{w}_{sj}^{(l)}$, $\mathbf{w}_{aj}^{(l)}$ and $\mathbf{w}_{vj}^{(l)}$ being the given beamforming vectors in the $l$-th iteration.
	
	Note that $\lvert {\mathbf{f}_{s,si}^H}\mathbf{w}_{sj} \rvert^2$, $\lvert {\mathbf{g}_{a,si}^H}\mathbf{w}_{aj} \rvert^2$ and $\lvert \mathbf{h}_{v,si}^H\mathbf{w}_{vj} \rvert^2$ are convex functions with respect to $\mathbf{w}_{sj}$, $\mathbf{w}_{aj}$ and $\mathbf{w}_{vj}$, respectively. Then, we approximate them with the first-order Taylor expansions as their lower bounds, which can be expressed as
\begin{equation}
	\begin{aligned}
		\lvert {\mathbf{f}_{s,si}^H}\mathbf{w}_{sj} \rvert^2& \! \geq \! 2\text{Re}\!\left\{\!\mathbf{w}_{sj}^{(l)H}{\mathbf{f}_{s,si}}{\mathbf{f}_{s,si}^H}\!\left(\!\mathbf{w}_{sj}\!-\!\mathbf{w}_{sj}^{(l)}\!\right)\!\right\}\!+\! \lvert {\mathbf{f}_{s,si}^H}\mathbf{w}_{sj}^{(l)} \rvert^2\\
		&\! \triangleq \mathbf{F}^{(l)}({\mathbf{f}_{s,si}},\mathbf{w}_{sj}).
	\end{aligned}
\end{equation}
	The approximation can be applied to other similar convex functions as $\lvert {\mathbf{g}_{a,si}^H}\mathbf{w}_{aj} \rvert^2 \geq \mathbf{F}^{(l)}({\mathbf{g}_{a,si}},\mathbf{w}_{aj})$, $\lvert \mathbf{h}_{v,si}^H\mathbf{w}_{vj} \rvert^2 \geq \mathbf{F}^{(l)}(\mathbf{h}_{v,si},\mathbf{w}_{vj})$ and so forth.
	
	Thereby, (\ref{opProb1}{f}) is transformed into a standard convex constraint. For simplicity, we introduce a slack variable set $\mathcal{S}=\{S_{vj}^{v,i}\triangleq \lvert \mathbf{h}_{v,i}^H \mathbf{w}_{vj} \rvert^2,
	S_{aj}^{a,i}\triangleq \lvert {\mathbf{g}_{a,i}^H} \mathbf{w}_{aj} \rvert^2 ,
	S_{sj}^{s,i}\triangleq \lvert {\mathbf{f}_{s,i}^H} \mathbf{w}_{sj} \rvert^2,
	S_{bv}^{b,X} \triangleq \lVert \mathbf{H}_{b,X}^H \mathbf{w}_{bv} \rVert^2,
	S_{ba}^{b,X} \triangleq \lVert \mathbf{H}_{b,X}^H \mathbf{w}_{ba} \rVert^2,
	S_{bi}^{b,X} \triangleq \lVert \mathbf{H}_{b,X}^H \mathbf{w}_{bi} \rVert^2,\forall i,j\}$, where $X \in \{v,a,bi\}$. Then, the same idea can be applied to (\ref{opProb1}{a})-(\ref{opProb1}{f}), which are transformed as
	\begin{align}
		\label{AppR_ba} &\sum\limits_{j = 1}^{{M_a}} {{r_{aj}}} \le {\log _2}\left( {S_{ba}^{b,a} + T_{ba} + \sigma _{ba}^2} \right) - \tilde R_{ba}^{\mathrm{up}}, \\
		\label{AppR_bv} &\sum\limits_{j = 1}^{{M_v}} {{r_{vj}}} \le {\log _2}\left( {S_{bv}^{b,v} + T_{bv}  + \sigma _{bv}^2} \right) - \tilde R_{bv}^{\mathrm{up}}, \\
		\label{AppR_ai} &\eta  \le {r_{ai}} \le {\log _2}\left( S_{ai}^{a,ai}  + T_{ai}+ \sigma _{ai}^2 \right) - \tilde R_{ai}^{\mathrm{up}}, \forall ai\\
		\label{AppR_vi} &\eta  \le {r_{vi}} \le {\log _2}\left( S_{vi}^{v,vi}  + T_{vi}+ \sigma _{vi}^2 \right) - \tilde R_{vi}^{\mathrm{up}}, \forall vi\\
		\label{AppR_bi} &\eta  \le {\log _2}\left( S_{bi}^{b,bi}  + T_{bi} + \sigma _{bi}^2 \right) - \tilde R_{bi}^{\mathrm{up}}, \forall bi\\
		\label{AppR_si} &\eta  \le {\log _2}\left( {S_{si}^{s,si} + T_{si}  + \sigma _{si}^2} \right) - \tilde R_{si}^{\mathrm{up}}, \forall si
	\end{align}
	 where $T_{ba}\!\!=\!\! S_{bv}^{b,a}+\sum_{j=1}^{M_b}S_{bj}^{b,a}$, $T_{bv}\!\!=\!\! S_{ba}^{b,v}+\sum_{j=1}^{M_b}S_{bj}^{b,v}, \\
	 T_{ai}\, = \, \sum_{j=1,j \neq i}^{M_a} S_{aj}^{a,ai}\, +\, \sum_{j=1}^{M_v}S_{vj}^{v,ai}\, +\, \sum_{j=1}^{M_s}S_{sj}^{s,ai}, \\
	 T_{vi}\,\, = \,\, \sum_{j=1,j \neq i}^{M_v} S_{vj}^{v,vi}\, +\, \sum_{j=1}^{M_a}S_{aj}^{a,vi}\, +\, \sum_{j=1}^{M_s}S_{sj}^{s,vi}\,\,, \\
	 T_{si}\, = \, \sum_{j=1,j \neq i}^{M_s} S_{sj}^{s,si}\, +\, \sum_{j=1}^{M_v}S_{vj}^{v,si}\, +\, \sum_{j=1}^{M_a}S_{aj}^{a,si}\, \text{and} \\
	 T_{bi}=\sum_{j=1,j \neq i}^{M_b} S_{bj}^{b,bi} + S_{bv}^{b,bi} + S_{ba}^{b,bi}+ \sum_{j=1}^{M_v}S_{vj}^{v,bi} + \sum_{j=1}^{M_a}S_{aj}^{a,bi}+\sum_{j=1}^{M_s}S_{sj}^{s,bi}$.
%	\begin{align}
%		T_{ba}&=S_{bv}^{b,a}+\sum_{j=1}^{M_b}S_{bj}^{b,a},\\
%		T_{bv}&=S_{ba}^{b,v}+\sum_{j=1}^{M_b}S_{bj}^{b,v},\\
%		T_{ai}&=\!\!\! \sum_{j=1,j \neq i}^{M_a}\!\!\! S_{aj}^{a,ai}+\sum_{j=1}^{M_v}S_{vj}^{v,ai}+\sum_{j=1}^{M_s}S_{sj}^{s,ai},\\
%		T_{vi}&=\!\!\! \sum_{j=1,j \neq i}^{M_v}\!\!\! S_{vj}^{v,vi} + \sum_{j=1}^{M_a}S_{aj}^{a,vi} + \sum_{j=1}^{M_s}S_{sj}^{s,vi},\\
%		T_{bi}&=\!\!\! \sum_{j=1,j\neq i}^{M_b}\!\!\! S_{bj}^{b,bi} + S_{bv}^{b,bi} + S_{ba}^{b,bi}\nonumber\\
%			& + \sum_{j=1}^{M_v}S_{vj}^{v,bi} + \sum_{j=1}^{M_a}S_{aj}^{a,bi}+\sum_{j=1}^{M_s}S_{sj}^{s,bi},\\
%		T_{si}&=\!\!\! \sum_{j=1,j\neq i}^{M_s}\!\!\! S_{sj}^{s,si}+\sum_{j=1}^{M_v}S_{vj}^{v,si}+\sum_{j=1}^{M_a}S_{aj}^{a,si}.
%	\end{align}
	
	 Finally, the beamforming optimization sub-problem can be expressed as
	\begin{align}
		\max \limits_{\mathcal{W},\mathcal{S},\mathcal{R}} \ &\eta \label{opProb2}\\
		\textrm{s.t.} &S_{vj}^{v,i} \le \mathbf{F}^{(l)}(\mathbf{h}_{v,i},\mathbf{w}_{vj}), \forall i \in \mathcal{I}\tag{\ref{opProb2}{a}}\\
		&S_{aj}^{a,i} \le \mathbf{F}^{(l)}({\mathbf{g}_{a,i}},\mathbf{w}_{aj}), \forall i \in \mathcal{I}\tag{\ref{opProb2}{b}}\\
		&S_{sj}^{s,i} \le \mathbf{F}^{(l)}({\mathbf{f}_{s,i}},\mathbf{w}_{sj}), \forall i \in \mathcal{I}\tag{\ref{opProb2}{c}}\\
		&S_{ba}^{b,X} \le \mathbf{F}^{(l)}(\mathbf{H}_{b,X},\mathbf{w}_{ba}), \forall X\tag{\ref{opProb2}{d}}
	\end{align}
	\begin{align}
		&S_{bv}^{b,X} \le \mathbf{F}^{(l)}(\mathbf{H}_{b,X},\mathbf{w}_{bv}), \forall X\tag{\ref{opProb2}{e}}\\
		&S_{bi}^{b,X} \le \mathbf{F}^{(l)}(\mathbf{h}_{b,X},\mathbf{w}_{bi}), \forall X, bi \in \mathcal{BI} \tag{\ref{opProb2}{f}}\\
		&(\ref{opProb}{\mathrm{b}})-(\ref{opProb}{\mathrm{e}}), (\ref{AppR_ba})-(\ref{AppR_si}).\nonumber
	\end{align}
	
	Since the problem (\ref{opProb2}) is convex, we can efficiently solve it with interior-point method \cite{Convex_Optimization}. It is worth pointing out that due to the DC programming technique, the solution of (\ref{opProb2}) is merely an approximate to the solution of original beamforming optimization sub-problem.
	\setcounter{TempEqCnt}{\value{equation}}
	\setcounter{equation}{61}
	\begin{figure*}[hb]
		\hrulefill
		\begin{equation}
			\begin{aligned}
				R_{si} =\hat{R}_{si}-\log_2\left(\sum_{j=1,j \neq i}^{M_s}\!\!\! \lvert {\mathbf{f}_{s,si}^H}\mathbf{w}_{sj} \rvert^2 + \sum_{j = 1}^{M_v}\frac{\lvert \overline{\mathbf{h}}_{v,si}^H\mathbf{w}_{vj}\rvert^2}{U_{v,si}} + \sum_{j = 1}^{M_a}\frac{\lvert\overline{{\mathbf{g}}}_{a,si}^H\mathbf{w}_{aj} \rvert^2}{L_{a,si}} + \sigma_{si}^2\right),
				\label{diffR_si2}
			\end{aligned}
		\end{equation}
		where $\hat R_{si}$ is given by
		\begin{equation}
			{\hat R_{si}} = \log_2\left(\sum_{j = 1}^{M_s}\lvert {\mathbf{f}_{s,si}^H}\mathbf{w}_{sj} \rvert^2 + \sum_{j = 1}^{M_v}\frac{\lvert \overline{\mathbf{h}}_{v,si}^H\mathbf{w}_{vj}\rvert^2}{U_{v,si}} + \sum_{j = 1}^{M_a}\frac{\lvert\overline{{\mathbf{g}}}_{a,si}^H\mathbf{w}_{aj} \rvert^2}{L_{a,si}} + \sigma_{si}^2\right),
		\end{equation}
	\end{figure*}
	\setcounter{equation}{\value{TempEqCnt}}
	\subsubsection{Trajectory Optimization with Fixed Beamforming}
	\
	\newline
	\indent In this sub-problem, we optimize trajectory $\mathcal{Q}$ with the fixed beamforming $\mathcal{W}$. In this case, (\ref{USVsafeObC.}), (\ref{USVsafeShC.}) and (\ref{opProb1}{a})-(\ref{opProb1}{f}) are non-convex constraints with respect to $\mathcal{Q}$.

	Firstly, we deal with constraints (\ref{USVsafeObC.}) and (\ref{USVsafeShC.}). Since $\lVert \mathbf{q}_v-\mathbf{o}_k \rVert^2$ and $\lVert \mathbf{q}_v-\mathbf{q}_i \rVert^2$ are convex functions with respect to $\mathbf{q}_v$, we approximate them with their first-order Taylor expansions and obtain the following inequalities
	\begin{align}
		\label{convUSVsafeObC.} \lVert \mathbf{q}_v-\mathbf{o}_k \rVert^2 \! &\geq \! 2{{\left( {{\bf{q}}_v^{\left( l \right)} - {{\bf{o}}_k}} \right)}^T}\!\!\left( {{{\bf{q}}_v} - {\bf{q}}_v^{\left( l \right)}} \right)\! +\! {{\left\| {{\bf{q}}_v^{\left( l \right)} - {{\bf{o}}_k}}\right\|}^2},\forall k\\
		\label{convUSVsafeShC.} \lVert \mathbf{q}_v-\mathbf{q}_i \rVert^2 \! &\geq \! 2{{\left( {{\bf{q}}_v^{\left( l \right)} - {{\bf{q}}_i}} \right)}^T}\!\! \left( {{{\bf{q}}_v} - {\bf{q}}_v^{\left( l \right)}} \right)\! +\! {{\left\| {{\bf{q}}_v^{\left( l \right)} - {{\bf{q}}_i}}\right\|}^2},\forall i.
	\end{align}
	
	For transmission rates, variables $\mathbf{q}_v$ and $\mathbf{q}_a$ are contained in the link distance expressions. From channel models, we can extract matrix components, which are functionally independent of $\mathbf{q}_v$ and $\mathbf{q}_a$, into $\overline{\mathbf{h}}_{v,i}$, $\overline{\mathbf{g}}_{a,i}$, $\overline{\mathbf{H}}_{b,v}$ and $\overline{{\mathbf{G}}}_{b,a},$ and obtain
	\begin{align}
		{{\bf{h}}_{v,i}} &\triangleq \frac{{{{\overline {\bf{h}} }_{v,i}}}}{{{{\left\| {{{\bf{q}}_v}- {{\bf{q}}_i}} \right\|}^2} + {{\left( {{h_v} - {h_i}} \right)}^2}}},\\
		{{\bf{g}}_{a,i}}&\triangleq \frac{{{{\overline {\bf{g}} }_{a,i}}}}{{\sqrt {{{\left\| {{{\bf{q}}_a} - {{\bf{q}}_i}} \right\|}^2} + {{\left( {{h_a} - {h_i}} \right)}^2}} }},\\
		{{\bf{H}}_{b,v}}&\triangleq \frac{{{{\overline {\bf{H}} }_{b,v}}}}{{{{\left\| {{{\bf{q}}_b}- {{\bf{q}}_v}} \right\|}^2} + {{\left( {{h_b} - {h_v}} \right)}^2}}}, \\
		{{\bf{G}}_{b,a}}&\triangleq \frac{{{{\overline {\bf{G}} }_{b,a}}}}{{\sqrt {{{\left\| {{{\bf{q}}_b} - {{\bf{q}}_a}} \right\|}^2} + {{\left( {{h_b} - {h_a}} \right)}^2}} }}.
	\end{align}
	Since these matrices are independent of $\mathbf{q}_v$ and $\mathbf{q}_a$, $\overline{\mathbf{h}}_{v,i}$, $\overline{\mathbf{g}}_{a,i}$, $\overline{\mathbf{H}}_{b,v}$ and $\overline{\mathbf{G}}_{b,a}$ can be regarded as given matrices. Moreover, $\mathbf{h}_{b,i}$ and ${\mathbf{f}_{s,i}}$ are also unrelated to $\mathbf{q}_v$ and $\mathbf{q}_a$, which are similarly viewed as given matrices.
	
	By defining ${U_{v,si}} = {[ {{{\left\| {{{\bf{q}}_v} - {{\bf{q}}_{si}}} \right\|}^2} + {{\left( {{h_v} - {h_i}} \right)}^2}} ]^2}$ and ${L_{a,si}} = {\left\| {{{\bf{q}}_a} - {{\bf{q}}_{si}}} \right\|^2} + {\left( {{h_a} - {h_i}} \right)^2}$, we can formulate $R_{si}$ into the difference of two logarithmic functions as (\ref{diffR_si2}).
	From Hessian matrix, it can be verified that (\ref{diffR_si2}) is the difference of two jointly convex functions with respect to $U_{v,si}$ and $L_{a,si}$. Then, DC programming is utilized by defining $D_{si}^{(l)}=\sum_{j = 1}^{M_s}\lvert {\mathbf{f}_{s,si}^{H}}\mathbf{w}_{sj} \rvert^2 + \frac{\sum_{j = 1}^{M_v}\lvert \overline{\mathbf{h}}_{v,si}^H\mathbf{w}_{vj}\rvert^2}{U_{v,si}^{(l)}} + \frac{\sum_{j = 1}^{M_a}\lvert\overline{\mathbf{g}}_{a,si}^H\mathbf{w}_{aj} \rvert^2}{L_{a,si}^{(l)}}$, where $U_{v,si}^{(l)}=[\lVert \mathbf{q}_v^{(l)}-\mathbf{q}_{si} \rVert^2 + (h_v-h_i)^2]^2$ and $L_{a,si}^{(l)}=\lVert \mathbf{q}_a^{(l)}-\mathbf{q}_{si} \rVert^2 + (h_a-h_i)^2$ with $\mathbf{q}_v^{(l)}$ and $\mathbf{q}_a^{(l)}$ being the given trajectory at $l$-th iteration. Therefore, we can obtain
	 \setcounter{equation}{63}
	\begin{equation}
		\begin{aligned}
			\hat{R}_{si}&\!\geq\! -B_{si,v}^{(l)}(U_{v,si}\!-\!U_{v,si}^{(l)})\!-\!B_{si,a}^{(l)}(L_{a,si}\!-\!L_{a,si}^{(l)})\\
			&\!+\log_{2}(D_{si}^{(l)}+\sigma_{si}^2)\!\triangleq \hat{R}_{si}^{\mathrm{lb}},
		\end{aligned}
		\label{Rlb_si}
	\end{equation}
	where $B_{si,v}^{(l)}$ and $B_{si,a}^{(l)}$ represent the first-order derivatives of $U_{v,si}$ and $L_{a,si}$ respectively, which are given by
	\begin{equation}
		B_{si,v}^{(l)}=\frac{\sum_{j=1}^{M_v}\lvert \overline{\mathbf{h}}_{v,si}^H\mathbf{w}_{vj} \rvert^2\log_2e}{\left(U_{v,si}^{(l)}\right)^2\left(D_{s,si}^{(l)}+\sigma_{si}^2\right)},
	\end{equation}
	\begin{equation}
		B_{si,a}^{(l)}=\frac{\sum_{j=1}^{M_a}\lvert \overline{\mathbf{g}}_{a,si}^H\mathbf{w}_{aj} \rvert^2\log_2e}{\left(L_{a,si}^{(l)}\right)^2\left(D_{s,si}^{(l)}+\sigma_{si}^2\right)}.
	\end{equation}
	
	Note that $U_{v,si}$ and $L_{a,si}$ are convex functions with respect to $\mathbf{q}_v$ and $\mathbf{q}_a$, respectively. Similarly, we can approximate them with their first order Taylor expansions as
	\begin{align}
		\label{K_vi} U_{v,si} &\geq U_{v,si}^{(l)}+4\sqrt{U_{v,si}^{(l)}}\left(\mathbf{q}_v^{(l)}-\mathbf{q}_{si}\right)^T\left(\mathbf{q}_v-\mathbf{q}_v^{(l)}\right),\\
		\label{L_ai} L_{a,si} &\geq L_{a,si}^{(l)}+2\left(\mathbf{q}_a^{(l)}-\mathbf{q}_{si}\right)^T\left(\mathbf{q}_a-\mathbf{q}_a^{(l)}\right).
	\end{align}
	
	Therefore, we turn (\ref{opProb1}{f}) into a standard convex constraint. For simplicity, we introduce slack variable sets $\mathcal{U}=\{u_{v,i} \triangleq U_{v,i}, u_{b,v} \triangleq U_{b,v},\forall i\}$ and $\mathcal{L}=\{l_{a,i} \triangleq L_{a,i}, l_{b,a} \triangleq L_{b,a},\forall i\}$. Then, the same idea can be applied to (\ref{opProb1}{a})-(\ref{opProb1}{f}), which are expressed as
	\begin{align}
		\label{AppR_ba2} &\sum_{j=1}^{M_a}r_{aj} \leq \hat{R}_{ba}^{\mathrm{lb}}
		-\log_2\left(D_{ba}-\frac{\lVert \overline{\mathbf{G}}_{b,a}^H\mathbf{w}_{ba} \rVert^2}{l_{b,a}}+\sigma_{ba}^2\right),\\
		\label{AppR_bv2} &\sum_{j=1}^{M_v}r_{vj} \leq \hat{R}_{bv}^{\mathrm{lb}}
		-\log_2\left(D_{bv}-\frac{\lVert \overline{\mathbf{H}}_{b,v}^H\mathbf{w}_{bv} \rVert^2}{u_{b,v}}+\sigma_{bv}^2\right),\\
		\label{AppR_ai2} &\eta \leq r_{ai} \leq \hat{R}_{ai}^{\mathrm{lb}}
		-\log_2\left(\!D_{ai}\!-\!\frac{\lvert \overline{\mathbf{g}}_{a,ai}^H\mathbf{w}_{ai} \rvert^2}{l_{a,ai}}\!+\!\sigma_{ai}^2\!\right), \forall ai\\
		\label{AppR_vi2} &\eta \leq r_{vi} \leq \hat{R}_{vi}^{\mathrm{lb}}
		-\log_2\left(\! D_{vi}\!-\!\frac{\lvert \overline{\mathbf{h}}_{v,vi}^H\mathbf{w}_{vi} \rvert^2}{u_{v,vi}}\!+\!\sigma_{vi}^2 \!\right), \forall vi\\
		\label{AppR_bi2} &\eta \leq \hat{R}_{bi}^{\mathrm{lb}}
		-\log_2\left(D_{bi}-\lvert \mathbf{h}_{b,bi}^H\mathbf{w}_{bi} \rvert^2+\sigma_{bi}^2\right), \forall bi
	\end{align}
	\begin{align}
		\label{AppR_si2}\!\!\!\!\!\!\!\! &\eta \leq \hat{R}_{si}^{\mathrm{lb}}
		-\log_2\left( D_{si}-\lvert {\mathbf{f}_{s,si}^H}\mathbf{w}_{si} \rvert^2+\sigma_{si}^2 \right), \forall si
	\end{align}
	where $D_{ba}\!=\!\dfrac{\lVert \overline{\mathbf{G}}_{b,a}^H\mathbf{w}_{ba} \rVert^2\!+\!\lVert \overline{\mathbf{G}}_{b,a}^H\mathbf{w}_{bv} \rVert^2\!+
		\!\sum_{j=1}^{M_b}\lVert \overline{\mathbf{G}}_{b,a}^H\mathbf{w}_{bj} \rVert^2}{l_{b,a}}, \\
	D_{bv}=\dfrac{\lVert \overline{\mathbf{H}}_{b,v}^H\mathbf{w}_{bv} \rVert^2+\lVert \overline{\mathbf{H}}_{b,v}^H\mathbf{w}_{ba} \rVert^2+\sum_{j=1}^{M_b}\lVert \overline{\mathbf{H}}_{b,v}^H\mathbf{w}_{bj} \rVert^2}{u_{b,v}}, \\
	D_{ai}\!=\!\dfrac{\sum_{j=1}^{M_a}\!\lvert \overline{\mathbf{g}}_{a,ai}^H\mathbf{w}_{aj} \rvert^2}{l_{a,ai}}\!+\!\dfrac{\sum_{j=1}^{M_v}\!\lvert \overline{\mathbf{h}}_{v,ai}^H\mathbf{w}_{vj} \rvert^2}{u_{v,ai}}\!+\!\sum_{j=1}^{M_s}\lvert {\mathbf{f}_{s,ai}^H}\mathbf{w}_{sj} \rvert^2\!\!,\\
	D_{vi}\!=\!\dfrac{\sum_{j=1}^{M_v}\!\lvert \overline{\mathbf{h}}_{v,vi}^H\mathbf{w}_{vj} \rvert^2}{u_{v,vi}}\!+\!\dfrac{\sum_{j=1}^{M_a}\!\lvert \overline{\mathbf{g}}_{a,vi}^H\mathbf{w}_{aj} \rvert^2}{l_{a,vi}}\!+\!\sum_{j=1}^{M_s}\lvert {\mathbf{f}_{s,vi}^H}\mathbf{w}_{sj} \rvert^2\!\!,\\
	D_{si}\!=\!\sum_{j = 1}^{M_s}\lvert {\mathbf{f}_{s,si}^{H}}\mathbf{w}_{sj} \rvert^2\! +\! \dfrac{\sum_{j = 1}^{M_v}\lvert \overline{\mathbf{h}}_{v,si}^H\mathbf{w}_{vj}\rvert^2}{u_{v,si}}\! +\! \dfrac{\sum_{j = 1}^{M_a}\lvert\overline{\mathbf{g}}_{a,si}^H\mathbf{w}_{aj} \rvert^2}{l_{a,si}}\!\!,\\ \text{and} \,
	D_{bi}=\sum_{j=1}^{M_b}\lvert \mathbf{h}_{b,bi}^H\mathbf{w}_{bj} \rvert^2+\lvert \mathbf{h}_{b,bi}^H\mathbf{w}_{bv} \rvert^2+\lvert \mathbf{h}_{b,bi}^H\mathbf{w}_{ba} \rvert^2+\!\dfrac{\sum_{j=1}^{M_a}\!\lvert \overline{\mathbf{g}}_{a,bi}^H\mathbf{w}_{aj} \rvert^2}{l_{a,bi}}\!+\!\dfrac{\sum_{j=1}^{M_v}\!\lvert \overline{\mathbf{h}}_{v,bi}^H\mathbf{w}_{vj} \rvert^2}{u_{v,bi}}\!+\!\sum_{j=1}^{M_s}\lvert {\mathbf{f}_{s,bi}^H}\mathbf{w}_{sj} \rvert^2,\\
	$
	Finally, the trajectory optimization sub-problem can be expressed as
	
	\begin{align}
		\max\limits_{\mathcal{Q},\mathcal{R},\mathcal{U},\mathcal{L}}& \ \eta\label{opProb3}\\
		\textrm{s.t.}
		&r_{\mathrm{ob},k}^2  \leq  2{{\left( {{\bf{q}}_v^{\left( l \right)} - {{\bf{o}}_k}} \right)}^T}\left( {{{\bf{q}}_v} - {\bf{q}}_v^{\left( l \right)}} \right)\! +\! {{\left\| {{\bf{q}}_v^{\left( l \right)} - {{\bf{o}}_k}} \right\|}^2}\nonumber \\
		&,\forall k,\tag{\ref{opProb3}{a}}\\
		&r_{\mathrm{sh}}^2  \leq  2{{\left( {{\bf{q}}_v^{\left( l \right)} - {{\bf{q}}_i}} \right)}^T}\left( {{{\bf{q}}_v} - {\bf{q}}_v^{\left( l \right)}} \right)\! +\! {{\left\| {{\bf{q}}_v^{\left( l \right)} - {{\bf{q}}_i}} \right\|}^2}\nonumber\\
		&,\forall i,\tag{\ref{opProb3}{b}}\\
		&u_{b,v} \leq U_{b,v}^{(l)} - \mathbf{b}_{b,v}^{(l)}\left(\mathbf{q}_v-\mathbf{q}_v^{(l)}\right),\tag{\ref{opProb3}{c}}\\
		&l_{b,a} \leq L_{b,a}^{(l)} - 2\left(\mathbf{q}_b-\mathbf{q}_a^{(l)}\right)^T\left(\mathbf{q}_a-\mathbf{q}_a^{(l)}\right), \tag{\ref{opProb3}{d}}\\
		&u_{v,i} \leq U_{v,i}^{(l)} + \mathbf{b}_{v,i}^{(l)}\left(\mathbf{q}_v-\mathbf{q}_v^{(l)}\right), \forall i \tag{\ref{opProb3}{e}}\\
		&l_{a,i} \leq L_{a,i}^{(l)} + 2\left(\mathbf{q}_a^{(l)}-\mathbf{q}_i\right)^T\left(\mathbf{q}_a-\mathbf{q}_a^{(l)}\right), \forall i \tag{\ref{opProb3}{f}}\\
		&(\ref{MaxUSVSpC.})-(\ref{USVangleInitC.}), (\ref{MaxUAVSpC.})-(\ref{UAVangleInitC.}), (\ref{opProb}{a}), (\ref{AppR_ba2})-(\ref{AppR_si2})\nonumber
	\end{align}
	where $\mathbf{b}_{b,v}^{(l)}=4\sqrt{U_{b,v}^{(l)}}\left(\mathbf{q}_b-\mathbf{q}_{v}^{(l)}\right)^T$ and $\mathbf{b}_{v,i}^{(l)}=4\sqrt{U_{v,i}^{(l)}}\left(\mathbf{q}_v^{(l)}-\mathbf{q}_{i}\right)^T$.
	Problem (\ref{opProb3}) is now convex and can be easily solved by using the interior-point method. Similar to the problem (\ref{opProb2}), the optimal solution of (\ref{opProb3}) is an approximate to the solution of original trajectory optimization sub-problem.
	\subsubsection{Overall Algorithm}
	\
	\newline
	\indent Based on the above two sub-problems, we derive an overall algorithm as shown in \textbf{Algorithm 1}. In the $l$-th iteration, $\mathcal{W}$ is optimized by solving the problem (\ref{opProb2}) with given $\mathcal{Q}^{(l)}$ and $\mathcal{W}^{(l)}$. Denoting the solution as $\mathcal{W}^{(l+1)}$, $\mathcal{Q}$ is then optimized by solving problem (\ref{opProb3}) based on $\mathcal{W}^{(l+1)}$ and $\mathcal{Q}^{(l)}$ with the output $\mathcal{Q}^{(l+1)}$. The result $\{\mathcal{W}^{(l+1)},\mathcal{Q}^{(l+1)}\}$ is used as input in the $l+1$-th iteration.
	The algorithm ceases when the difference of objective function between iterations is below a given threshold $\Psi$ and output $\{\mathcal{W}^{*},\mathcal{Q}^{*}\}$ as local optimal beamforming and trajectory at time slot $n$.
	Above process repeats in each time slot and forms the cooperative strategies of the entire system during the whole transmission period.
	The algorithm can be solved in real time, based on the real-time state of channels and users obtained by Automatic Identification System and channel estimation techniques in each time slot.

	\begin{algorithm}
	\renewcommand{\algorithmicrequire}{\textbf{Input:}}
	\renewcommand{\algorithmicensure}{\textbf{Output:}} %替换关键词
	
	\caption{Joint Beamforming and Trajectory Optimization}
	\label{algorithm1}
	\begin{algorithmic}[1]
		\REQUIRE $\{\mathbf{q}_i[n]\}_{n=1}^{N}$\ and initial location $\mathcal{Q}[0]$
		\ENSURE $\{\mathcal{Q}^{*}[n],\mathcal{W}^{*}[n]\}_{n=1}^{N}$
		
		\STATE Initialize system parameters, set time slot $n=1$.
		
		\WHILE{$n \leq N$}
			\STATE Initialize iteration index $l=0$, $\mathcal{Q}^{(0)}[n]$ and $\mathcal{W}^{(0)}[n]$.
			\REPEAT
				\STATE	Given $\mathcal{Q}^{(l)}[n]$, obtain $\mathcal{W}^{(l+1)}[n]$ by solving (\ref{opProb2}) via the interior-point method.
				
				\STATE	Given $\mathcal{W}^{(l+1)}[n]$, obtain $\mathcal{Q}^{(l+1)}[n]$ by solving (\ref{opProb3}) via the interior-point method.
				
				\STATE	Set $l = l+1$.
			\UNTIL{{$\eta(\mathcal{Q}^{(l)},\mathcal{W}^{(l)}) - \eta(\mathcal{Q}^{(l-1)},\mathcal{W}^{(l-1)}) \leq \Psi$}}
			\STATE Store the output $\mathcal{W}^{*}[n]$ and $\mathcal{Q}^{*}[n]$ as beamforming and trajectory at time slot $n$.
			\STATE Set $n=n+1$.
		\ENDWHILE
	\end{algorithmic}
\end{algorithm}
		\begin{table}[t]
		\centering
		\caption{Simulation Parameter}
		\begin{tabular}{|c|c|}
			\hline
			Parameter                       																			& Value                         \\ \hline
			Carrier frequency                    																	& { 2GHz}                 	\\ \hline
			USV maximum speed $V_v^{\mathrm{max}}$                   					& 37.5m/s                 		\\ \hline
			USV maximum steering angle $\varphi_v^{\mathrm{max}}$					& $60^{\circ}$					\\ \hline
			UAV maximum speed $V_a^{\mathrm{max}}$                    				  & 50m/s                			 \\ \hline
			UAV maximum steering angle $\varphi_a^{\mathrm{max}}$					& $60^{\circ}$					\\ \hline
			Current velocity $\mathbf{v}_c$                       										& 5m/s                				\\ \hline
			Wind velocity $\mathbf{v}_w$																& 5m/s								\\ \hline
			Number of TBS antennas $N_b$                											  & 128                    				 \\ \hline
			Number of USV antennas $N_v$                											& 4                     				  \\ \hline
			Number of UAV antennas $N_a$                											& 4                     				 \\ \hline
			Number of satellite antennas $N_s$                										& 128                     				\\ \hline
			Number of TBS users $M_b$                  													& 2                       				  \\ \hline
			Number of USV users $M_v$                  											  		& 2                       				 \\ \hline
			Number of UAV users $M_a$                  													& 2                       				 \\ \hline
			Number of satellite users $M_s$                  											& 2                       					\\ \hline
			Effective antenna height of TBS $h_b$           										& 35m         							\\ \hline
			Effective antenna height of USV $h_v$            										& 15m         							\\ \hline
			Effective antenna height of UAV $h_a$            									  & 200m        						 \\ \hline
			Effective antenna height of users $h_i$            										& 10m         							\\ \hline
			Altitude of LEO satellite $h_s$           													 & 500km         						\\ \hline
			Transmit power of TBS $P_b$ 															  & 48.45dBm 							\\ \hline
			Transmit power of USV $P_v$ 															& 47dBm 								\\ \hline
			Transmit power of UAV $P_a$ 															& 47dBm 								\\ \hline
			Transmit power of satellite $P_s$ 														& 48.45dBm 								\\ \hline
			TBS antenna gain 												   								 & 33dBi 										\\ \hline
			USV antenna gain 												    							& 23dBi 										\\ \hline
			UAV antenna gain 												    							& 15dBi 										\\ \hline
			Satellite antenna gain 												  							& 55dBi 										\\ \hline
			Receive antenna gain																		  & 15dBi    										\\ \hline
			Rain fading mean $\mu_r$																  & -2.6dB 										\\ \hline
			Rain fading variance $\sigma_r^2$													 & 1.63dB 										\\ \hline
			Antenna spacing $b$                             											  & { 0.075m} 										\\ \hline
			Length of time slot $\Delta \tau$ 													   & 1s													\\ \hline
			Transmission period $T$																& 1000s													\\ \hline
			Rician factor $K$																				  & 20                       						\\ \hline
			Variance of AWGN $\sigma^2$     													& -110dB                        				\\ \hline
			Convergence threshold $\Psi$															& 0.001									 	\\ \hline
		\end{tabular}
		\label{table2}
	\end{table}
	\subsection{Algorithm Analysis}
	In this subsection, we analyze the convergence properties and computational complexity of our proposed algorithm.
	\subsubsection{Convergence Analysis}
	\
	\newline
	\indent	
	Let $\eta_{\mathrm{bmf}}^{\mathrm{lb},l}(\mathcal{Q}^{(l)},\mathcal{W}^{(l)})$ and $\eta_{\mathrm{trj}}^{\mathrm{lb},l}(\mathcal{Q}^{(l)},\mathcal{W}^{(l)})$ be the objective value by solving problems (\ref{opProb2}) and (\ref{opProb3}) at the given $\mathcal{Q}^{(l)}$ and $\mathcal{W}^{(l)}$, respectively. We also define $\eta(\mathcal{Q}^{(l)},\mathcal{W}^{(l)})$ as the objective value of problem (\ref{opProb}), which is denoted as the exactly optimal result since no approximation is introduced. Therefore, given $\mathcal{Q}^{(l)}$ and $\mathcal{W}^{(l)}$, we can obtain
	\begin{equation}
		\begin{aligned}
			\eta(\mathcal{Q}^{(l)},\mathcal{W}^{(l)})&\overset{\text{a}}{=}\eta_{\mathrm{bmf}}^{\mathrm{lb},l}(\mathcal{Q}^{(l)},\mathcal{W}^{(l)})\\
			&\overset{\text{b}}{\leq} \eta_{\mathrm{bmf}}^{\mathrm{lb},l}(\mathcal{Q}^{(l)},\mathcal{W}^{(l+1)})\\
			&\overset{\text{c}}{\leq} \eta(\mathcal{Q}^{(l)},\mathcal{W}^{(l+1)}),
		\end{aligned}
		\label{BmfConverge}
	\end{equation}
	where equation (a) is derived by the first-order Taylor expansion, inequality (b) can be obtained by solving the problem (\ref{opProb2}) with the solution $\mathcal{W}^{(l+1)}$, and inequality (c) can be verified since the objective of problem (\ref{opProb2}) is the lower bound of the optimal value of problem (\ref{opProb}) at $\mathcal{W}^{(l+1)}$. Therefore, expression (\ref{BmfConverge}) indicates that the objective value $\eta$ is non-decreasing after the beamforming optimization, namely, the beamforming is correctly optimized via step 5.
	
	Similarly, given $\mathcal{Q}^{(l)}$ and $\mathcal{W}^{(l+1)}$, we can obtain
	\begin{equation}
		\begin{aligned}
			\eta(\mathcal{Q}^{(l)},\mathcal{W}^{(l+1)})&\overset{\text{a}}{=}\eta_{\mathrm{trj}}^{\mathrm{lb},l}(\mathcal{Q}^{(l)},\mathcal{W}^{(l+1)})\\
			&\overset{\text{b}}{\leq} \eta_{\mathrm{trj}}^{\mathrm{lb},l}(\mathcal{Q}^{(l+1)},\mathcal{W}^{(l+1)})\\
			&\overset{\text{c}}{\leq} \eta(\mathcal{Q}^{(l+1)},\mathcal{W}^{(l+1)}).
		\end{aligned}
		\label{TrjConverge}
	\end{equation}
	With expressions (\ref{BmfConverge}) and (\ref{TrjConverge}), we can prove that the objective is always non-decreasing after each iteration. Due to transmit power constraints, the objective is evidently upper bounded by a finite value, thus $\eta$ should converge to a finite value.
	\subsubsection{Complexity Analysis}
	\
	\newline
	\indent Noticing that there are two loops nested in \textbf{Algorithm 1}, namely time loop with $N$ time slots and iteration loop until the objective converges. In each iteration, problem (\ref{opProb2}) in step 5 and problem (\ref{opProb3}) in step 6 are solved via the interior-point method. A custom-built interior-point solver can solve a convex problem with a complexity of $\mathcal{O}(n^{3.5}\log(1/\varepsilon))$, where $n$ denotes the problem scale and $\varepsilon$ denotes the error tolerance \cite{complexity}.
	
	Therefore, the complexity of \textbf{Algorithm 1} is given by $\mathcal{O}(N((N_bM_b+N_vM_v+N_aM_a+N_sM_s)^{3.5}+(M_a+M_v)^{3.5})\Gamma\log(1/\varepsilon))$, where $N$ denotes the number of time slots and $\Gamma$ denotes the average number of iterations required for convergence. The problem scale of (\ref{opProb2}) is considerably larger than (\ref{opProb3}), which suggests that the beamforming optimization is the dominant source of computational complexity.
	\section{SIMULATION RESULT}
	In this section, we provide extensive simulation results to validate the proposed algorithm and evaluate its performance.
	The simulation parameters are listed in Table \ref{table2}.
	\subsection{Trajectory Design}
	\begin{figure}[ht]
		\centering
		\subfigure[Scenario of separated and mobile users.]{\label{TraComImg}\includegraphics[width=0.8\columnwidth]{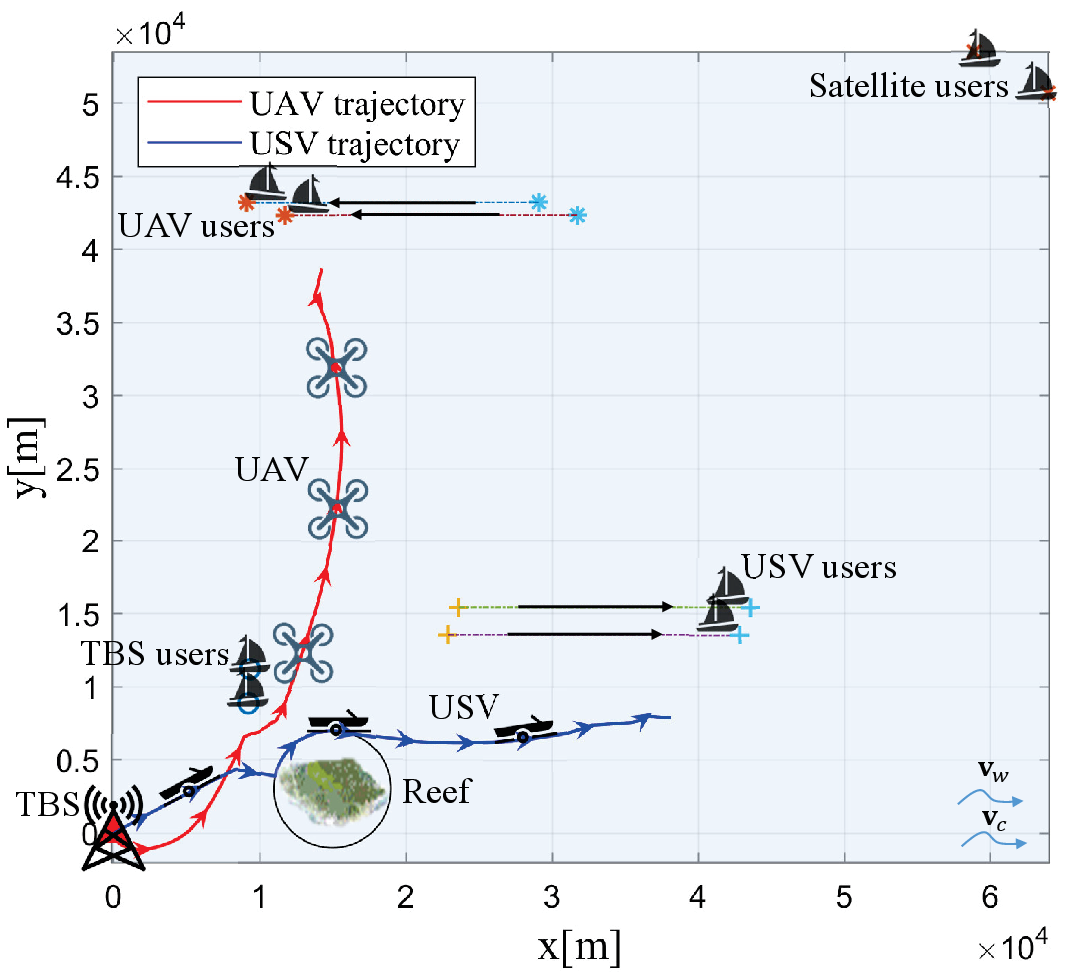}}\\
		\subfigure[Scenario of assembled and static users.]{\label{TraAssImg}\includegraphics[width=0.81\columnwidth]{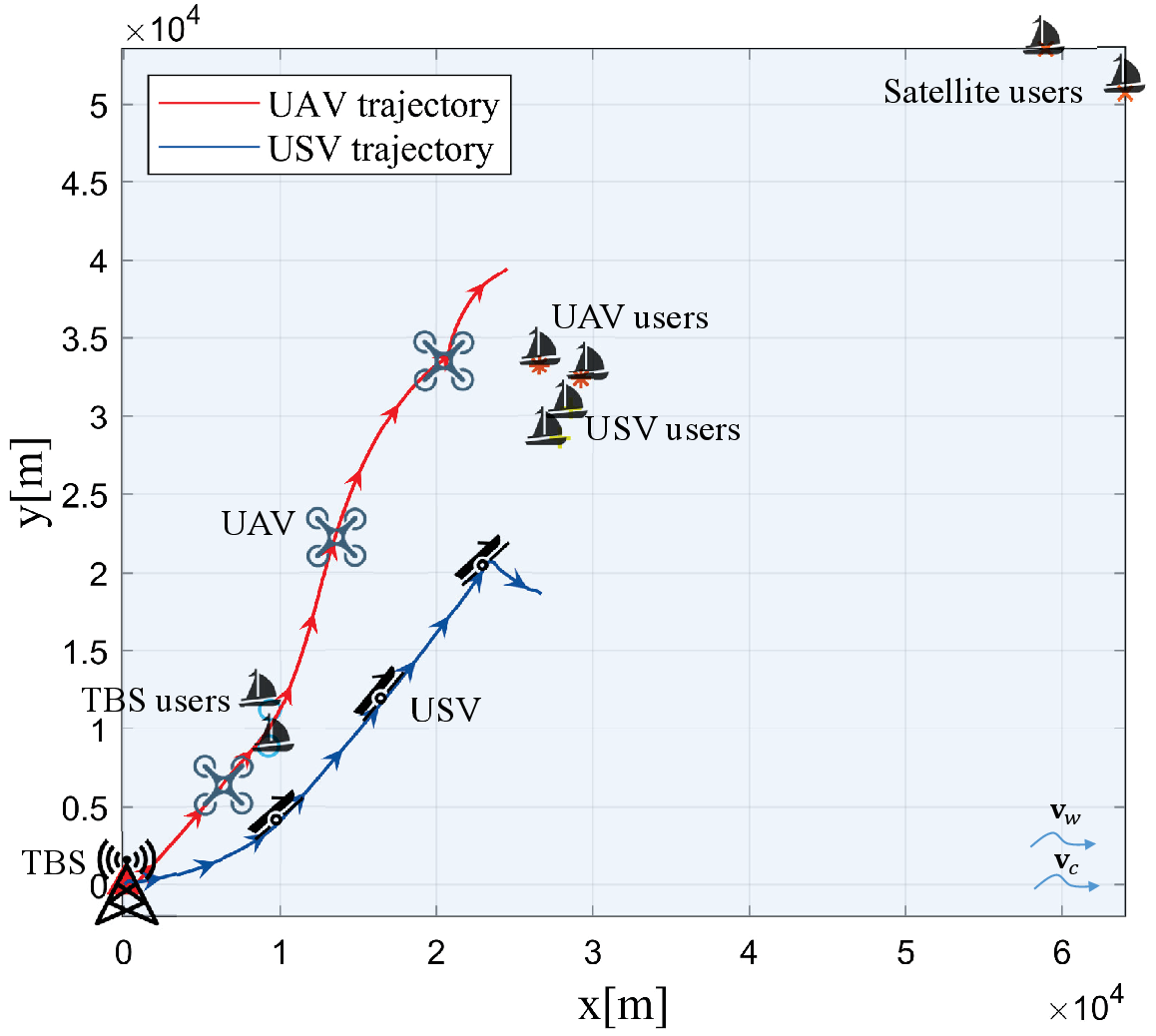}}
		\caption {The optimized trajectories of USV and UAV in different scenarios.}
		\label{Tra}
	\end{figure}

	We test the trajectory planning via deterministic simulation. Fig. \ref{TraComImg} reveals the optimized trajectories of USV and UAV. The USV's and UAV's initial locations are set to be $[0,0]^T$ and they set out simultaneously to provide on-demand services. The distribution of maritime users is arranged artificially to align with the regional division illustrated in Fig. \ref{SystemImg}. To better demonstrate the characteristics of trajectory planning, the USV and UAV users are positioned far apart and move in uniform linear motion at a speed of 20 m/s along the positive and negative x-axis directions, respectively. It is observed that the USV and UAV travel along the optimized trajectory approaching respective target users as close as possible. Moreover, the USV tends to sail around the reef due to the safe sailing constraints and the UAV tends to keep a distance from the TBS users to decrease the interference to them.
	
	Fig. \ref{TraAssImg} simulates the situation where static USV and UAV users are located near each other. Due to the close proximity between users, the beams transmitted by the USV would cause significant interference on the UAV users, and vice versa. Based on the performance fairness principle, \textbf{Algorithm 1} adopts a trade-off by maintaining a distance between both the UAV/USV and their users to ensure an overall improvement of system performance. The simulation reflects the unified design and fundamental trade-offs in the proposed architecture.
	
\begin{figure}[ht]
		\centering
		\subfigure[UAV users sail along the positive direction of x-axis.]{\label{TraCloImg}\includegraphics[width=0.8\columnwidth]{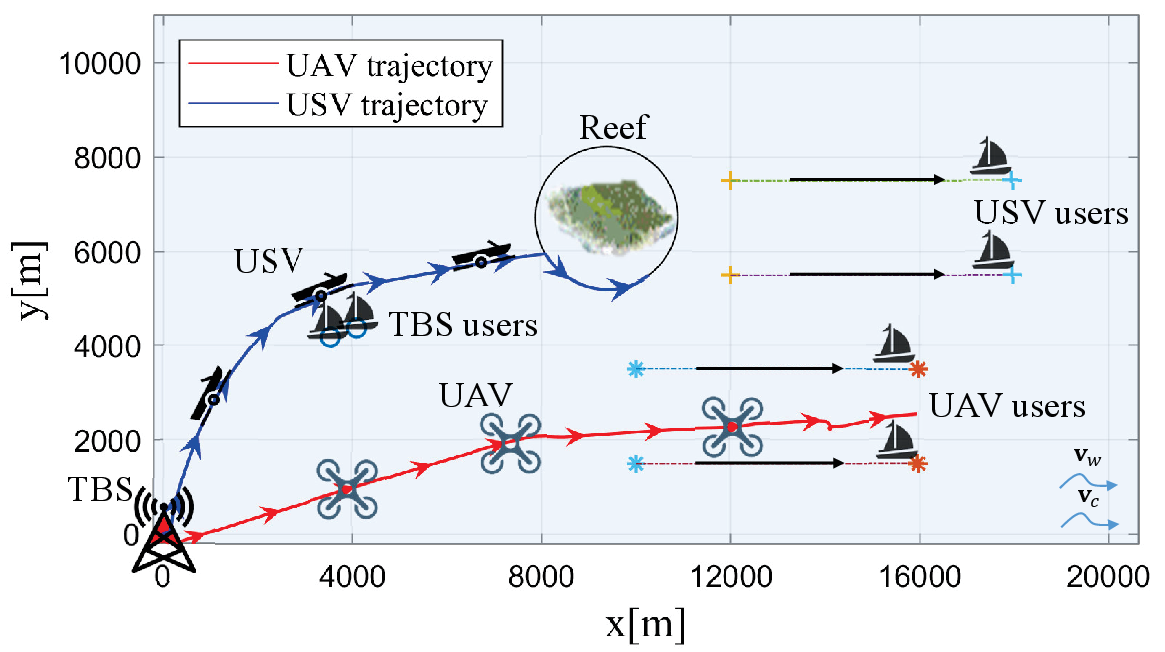}}\\
		\subfigure[UAV users sail along the negative direction of x-axis.]{\label{TraLonImg}\includegraphics[width=0.79\columnwidth]{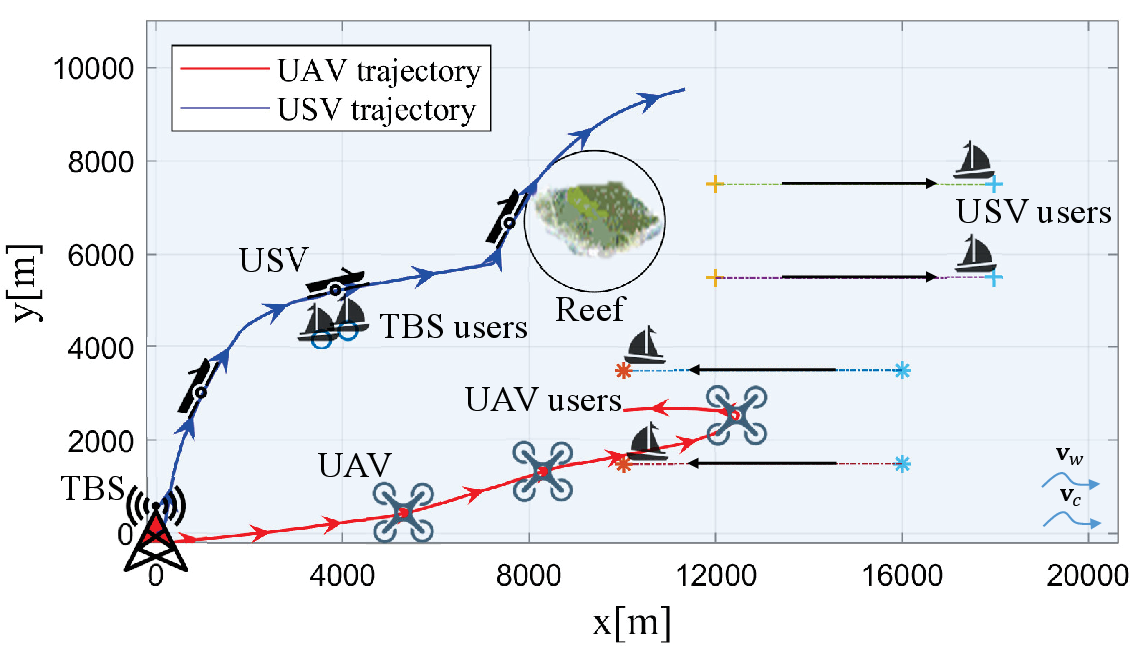}}
		\caption {The trajectory comparison about the impact of interference.}
		\label{TraComparison}
	\end{figure}

	To obtain insight into the impact of interference on trajectory, we conduct a pair of comparison tests as shown in Fig. \ref{TraComparison}, where the traveling direction of the UAV users is the only difference. Fig. \ref{TraCloImg} shows that the USV choose to sail along the bottom boundary of the reef to approach the USV fleet as quickly as possible. However, in Fig. \ref{TraLonImg}, the USV prefers a longer route around the top boundary of the reef to mitigate the interference to the coming UAV users, even though decreasing the transmission rates for USV users. The comparison indicates that the trajectory design is based on the comprehensive consideration of multiple users' benefits, forming a unified space-air-ground-sea integrated maritime communication system.

	\subsection{Performance Evaluation}	
\begin{figure}[t]
		\centering
		\includegraphics[width=0.8\columnwidth]{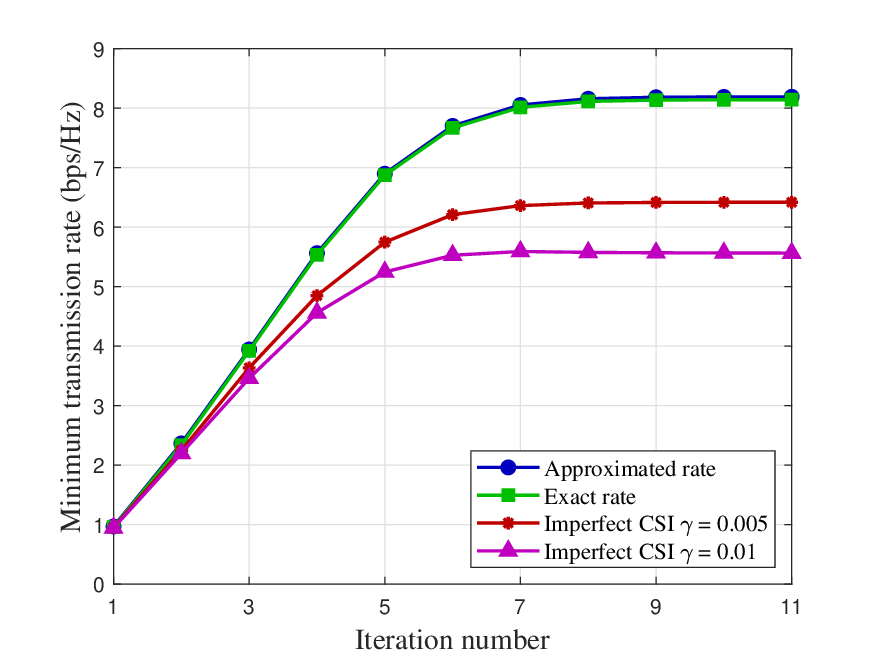}
		\caption{The convergence behaviors of minimum transmission rate for different states.}
		\label{GamConvImg}
	\end{figure}

\begin{figure}[t]
		\centering
		\includegraphics[width=0.8\columnwidth]{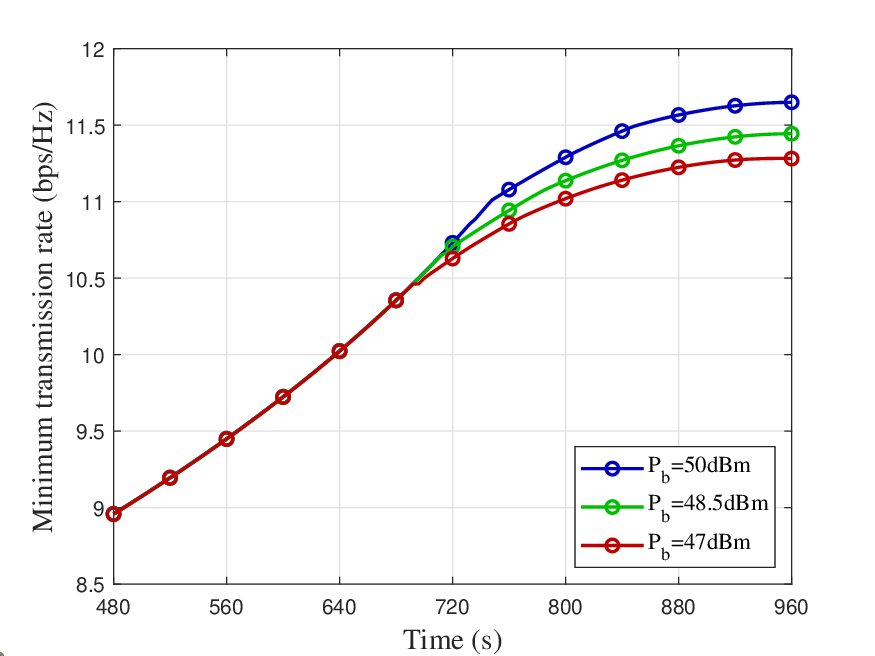}
		\caption{The minimum transmission rate at each time slot under different TBS transmit power.}
		\label{GamPBImg}
	\end{figure}

\begin{figure}[t]
		\centering
		\includegraphics[width=0.8\columnwidth]{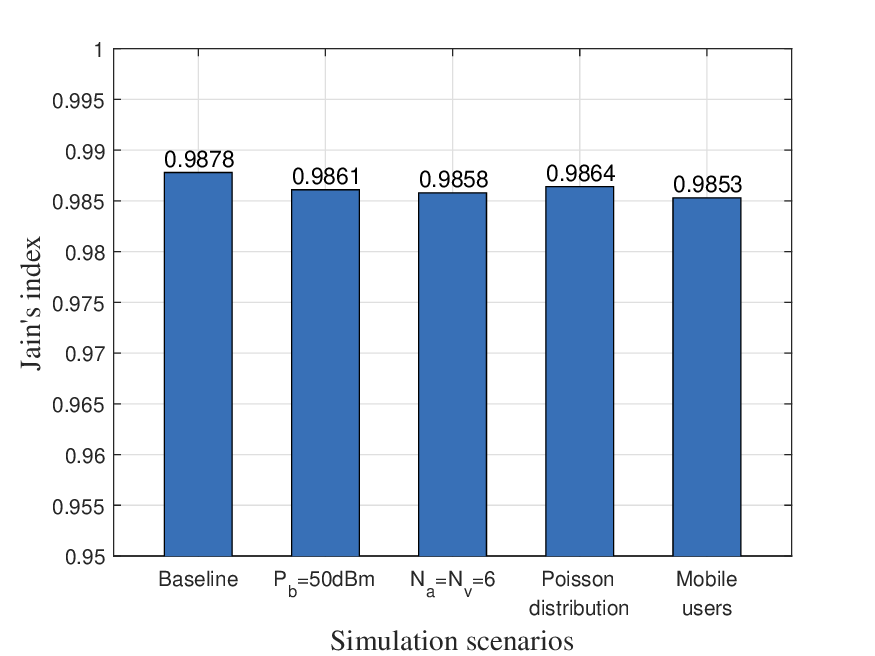}
		\caption{The Jain's index in different simulation scenarios.}
		\label{GamJain}
	\end{figure}

\begin{figure}[t]
		\centering
		\includegraphics[width=0.8\columnwidth]{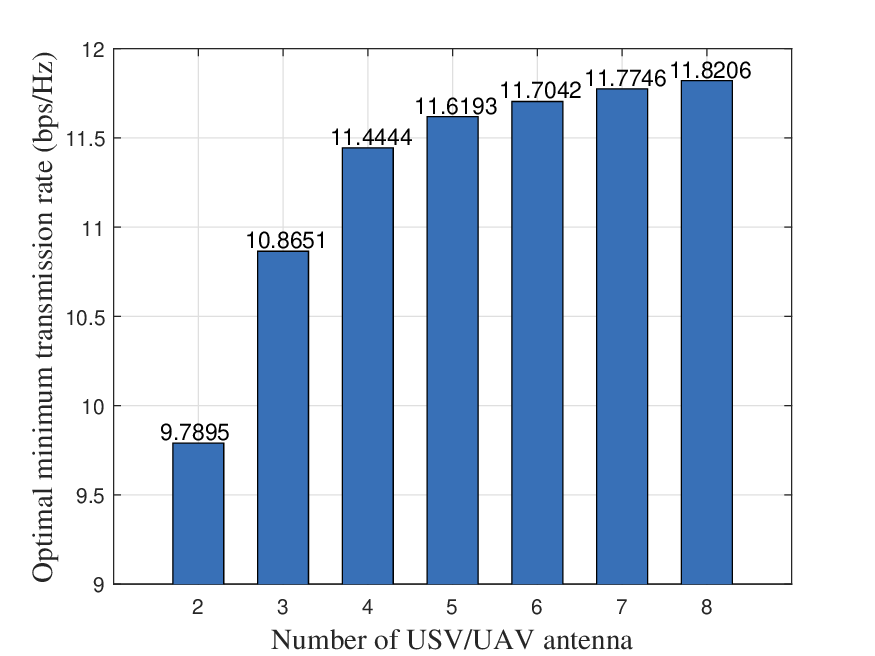}
		\caption{The minimum transmission rate at each time slot under different antenna numbers of the USV and UAV, where the antenna numbers of the USV and UAV are set to be equal.}
		\label{GamNANVImg}
	\end{figure}

\begin{figure}[t]
		\centering
		\includegraphics[width=0.8\columnwidth]{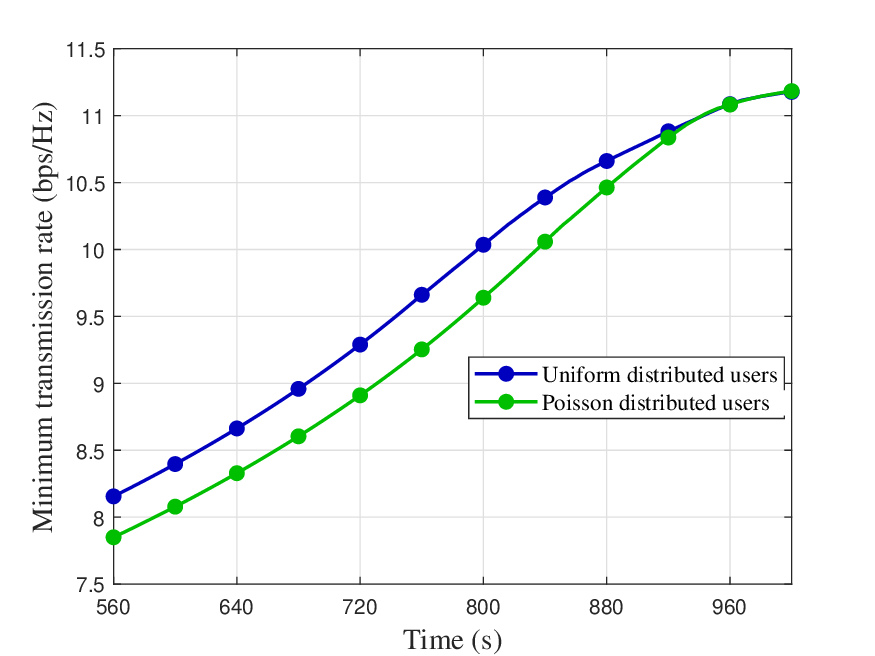}
		\caption{The minimum transmission rate at each time slot under different distribution of users.}
		\label{GamDisImg}
	\end{figure}

\begin{figure}[t]
		\centering
		\includegraphics[width=0.8\columnwidth]{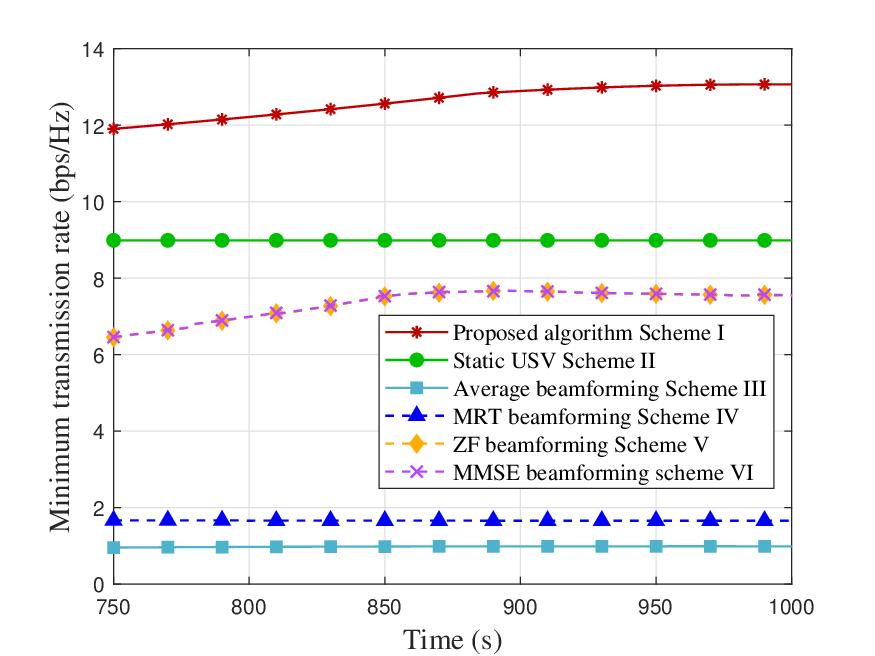}
		\caption{The minimum transmission rate at each time slot under different optimization schemes.}
		\label{GamAlgImg}
	\end{figure}

	We study the convergence behavior, the potential promoting factors and performance of the proposed algorithm in Fig. \ref{GamConvImg}--Fig. \ref{GamAlgImg}.
	These simulations use 100 runs to average the performance, realized by changing the seed of random number.
		
	Fig. \ref{GamConvImg} plots the convergence behavior for approximated rate and exact rate. It demonstrates that the proposed \textbf{Algorithm 1} converges in about 9 iterations in both approximated and exact cases. Hence, the proposed algorithm has a fast convergence speed. Additionally, the performance gap between exact rate and approximated rate is negligible, which indicates that the approximated two-ray channel effectively diminishes the sine function with trivial deviation. Furthermore, we study the impact of imperfect CSI by using a bounded model, which is suitable to capture the effect of instantaneous CSI variation at sea \cite{VIChannel9956800}. We define the channel uncertainty ratio as $\gamma = \frac{\delta}{\sqrt{\mathbb{E}\left\lbrace\lVert\hat{\mathbf{h}} \rVert^2 \right\rbrace }}$ with $\delta$ being the bound of CSI error and $\hat{\mathbf{h}}$ being the original CSI. It is seen that performance degradation brought by the imperfect CSI is significant and increases with the channel uncertainty ratio. To alleviate the impact of imperfect CSI, it makes sense to design robust beamforming schemes as done in \cite{satellite_terrestrial_integration,AIChannel8974403,VIChannel9956800}.
		
	Fig. \ref{GamPBImg} illustrates the impact of TBS transmit power on system performance under static user conditions. For a fixed TBS power, the rates increase over time due to the joint optimization of beamforming and trajectory, ultimately converging to an optimal value. The optimal value will increase with the increment of TBS power. Notably, the rates exhibit negligible variation with increased TBS power during the initial transmission phase. Consequently, rather than sustaining high transmit power throughout the entire transmission period, the TBS should progressively escalate its power over time to minimize energy consumption.
	
	
	Fig. \ref{GamJain} shows the result of Jain's indexes of bit rate allocation among users in different scenarios. It is a metric used to quantitatively measure the fairness of resource allocation among multiple users in a system \cite{Jain}. The baseline scenario is configured with transmit power of $P_b=48.5{\rm dBm}$, antenna number of $N_a=N_v=4$ and uniformly distributed static users. To test adaptability, additional scenarios are simulated under various configurations. Regardless of the variation, the index remains close to 1 (indicating perfect fairness). The results validate that the proposed algorithm is able to ensure equitable resource allocation under various conditions.
	
	Fig. \ref{GamNANVImg} reveals the optimal value of minimum transmission rates under various USV's and UAV's antenna number. Apparently, the transmission rate increases with the antenna number, and the rate improvement is gradually reducing. This is expected since the increase in antenna number brings more degrees of freedom (DOFs) for beamforming optimization to generate more effective beamforming vectors. Finally, the number of DOF provided by antenna number reaches saturation. Considering that the beamforming optimization contributes more computational complexity, the USV and UAV should be equipped with appropriate number of antennas to balance the complexity and performance.
	
	Fig. \ref{GamDisImg} shows the impact of user distribution of USV and UAV on the transmission rate. The simulation considers two distinct spatial distributions---uniform distribution and Poisson distribution---within the same areas. The result demonstrates that the spatial distribution of users affects the achievable transmission rate during service provisioning. However, the performance ultimately converges to similar optimal values under identical system parameters, highlighting the adaptability of the designed architecture to various user distributions.
	
	Finally, to investigate the performance of \textbf{Algorithm 1}, Fig. \ref{GamAlgImg} compares the following six schemes. Scheme \RMN{1}: All variables are jointly optimized by \textbf{Algorithm 1}; Scheme \RMN{2}: Optimized beamforming scheme but with static USV and UAV, which are located at the geometric center of the TBS and user fleets; Scheme \RMN{3}: Optimized trajectories with average beamforming; Scheme \RMN{4}: Optimized trajectories with maximum ratio transmission (MRT) beamforming \cite{VIChannel9956800};
	Scheme \RMN{5}: Optimized trajectories with zero forcing (ZF) beamforming; Scheme \RMN{6}: Optimized trajectories with minimum mean square error (MMSE) beamforming.
	
	Comparing Schemes \RMN{1}, \RMN{3}, \RMN{4}, \RMN{5} and \RMN{6}, the proposed algorithm achieves better performance than other beamforming schemes. This is because the average and MRT beamformings bring less DOFs and cannot fully eliminate complicated interference in the considered maritime system. Despite that the ZF and MMSE beamformings effectively mitigate co-channel interference, they exhibit limited capability in suppressing adjacent-channel interference. Note that the performance gap between ZF and MMSE schemes is negligible. It suggests that, as the system approaches its optimal state, adjacent-channel interference becomes the major factor degrading performance. This can be the reason why our proposed algorithm achieves better performance.
	Furthermore, the large gap between Scheme \RMN{2} and Scheme \RMN{4} indicates that an appropriate beamforming scheme is able to provide large performance gain in the considered maritime communication system.
	Overall, the comparison validates the effectiveness of the proposed algorithm in maritime communications.
	
	\section{CONCLUSION}
	In this paper, we proposed a novel space-air-ground-sea integrated maritime communication architecture combining TBS, USV, UAV and satellite to support seamless communication services. The channel model and signal model have been investigated and established. Accordingly, we proposed a performance optimization algorithm by jointly optimizing the trajectory and beamforming. Extensive simulation results have confirmed that the proposed algorithm was efficient to promote the system performance in space-air-ground-sea integrated maritime communications.
	In future work, we plan to discuss additional factors such as Doppler effect and imperfect CSI in maritime communications. Another possible extension of the research is the design of more robust and low-complexity optimization algorithm.

\end{document}